\documentclass[acmsmall]{acmart}

\usepackage{xcolor}
\usepackage{svg}
\usepackage{amsmath}
\usepackage{booktabs}
\usepackage{multirow}
\usepackage{array}  
\usepackage{graphicx}
\usepackage{alltt}
\usepackage{caption}
\usepackage{soul}
\usepackage{xspace}
\usepackage{enumitem}
\usepackage{tabularx}
\usepackage{siunitx}
\usepackage{pifont}
\usepackage{wrapfig}

\newcommand{\valid}{valid\xspace}
\newcommand{\invalid}{invalid\xspace}
\newcommand{\ckpt}{\textsf{ckpt}\xspace}

\newcommand{\isignpost}[1]{\textbf{#1}.}
\newcommand{\sys}{diffRL\xspace}

\definecolor{darkgreen}{rgb}{0.0, 0.5, 0.0}

\begin{document}

\title{Analyzing Symbolic Properties for DRL Agents in Systems and Networking}

\author{Mohammad Zangooei}
\affiliation{
  \institution{University of Waterloo}
  \city{Waterloo}
  \country{Canada}
}

\author{Jannis Weil}
\affiliation{
  \institution{Leibniz University Hannover}
  \city{Hannover}
  \country{Germany}
}

\author{Amr Rizk}
\affiliation{
  \institution{Leibniz University Hannover}
  \city{Hannover}
  \country{Germany}
}

\author{Mina Tahmasbi Arashloo}
\affiliation{
  \institution{University of Waterloo}
  \city{Waterloo}
  \country{Canada}
}

\author{Raouf Boutaba}
\affiliation{
  \institution{University of Waterloo}
  \city{Waterloo}
  \country{Canada}
}

\setcopyright{cc}
\setcctype{by}
\acmJournal{POMACS}
\acmYear{2026} \acmVolume{10} \acmNumber{2} \acmArticle{29}
\acmMonth{6} \acmDOI{10.1145/3805627}

\begin{abstract}

Deep reinforcement learning (DRL) has shown remarkable performance on complex control problems in systems and networking, including adaptive video streaming, wireless resource management, and congestion control. For safe deployment, however, it is critical to reason about how agents behave across the range of system states they may encounter in practice. Existing verification-based approaches in this domain primarily focus on \emph{point properties} -- properties defined around fixed input states -- which offer limited coverage and require substantial manual effort to identify relevant input-output pairs for analysis.

In this paper, we study \emph{symbolic properties} -- properties that specify expected behaviors over ranges of input states -- for DRL agents in systems and networking.
We present a generic formulation for symbolic properties, with monotonicity and robustness as concrete examples, and show how they can be analyzed using existing DNN verification engines. Our approach encodes symbolic properties as comparisons between related executions of the same policy and decomposes them into practically tractable sub-properties.
These techniques serve as practical enablers for applying existing verification tools to symbolic analysis.

Using our framework, \sys, we conduct an extensive empirical study across three representative DRL-based control systems -- adaptive video streaming, wireless resource allocation, and congestion control -- covering both discrete and continuous action spaces. Through these case studies, we analyze symbolic properties over broad input ranges, examine how property satisfaction evolves during training, study the impact of model size on verifiability, and compare multiple verification backends. Our results show that symbolic properties provide substantially broader coverage than point properties and can uncover non-obvious, operationally meaningful counterexamples, while also revealing practical solver trade-offs and limitations.

\end{abstract}

\maketitle

\section{Introduction} \label{sec:intro}

Deep Reinforcement Learning (DRL) has emerged as a promising approach for control problems in systems and networking, where decisions must be made in complex and highly dynamic environments. In these settings, DRL agents are typically embedded in control loops. The agent observes the system state -- represented as a feature vector of performance indicators such as throughput, latency, packet loss, or resource utilization -- and selects actions according to its learned policy.
These actions correspond to concrete numerical control decisions for resource allocation~\cite{zangooei2023flexible, park2024topfull, qiu2020firm, chinchali2018cellular}, traffic engineering and path selection~\cite{chen2018auto, xu2023teal, gui2024redte}, job scheduling~\cite{mao2019learning, mao2016resource, jog2021one, ko2024edgeric}, and tuning streaming video bitrate~\cite{mao2017neural, jia2023rdladder, jia2024dancing} or congestion window sizes~\cite{yan2021acc, tessler2022reinforcement, abbasloo2020classic, ma2022multi}.

To safely and effectively integrate DRL agents into such control loops, it is essential to reason about how ``well-behaved'' these agents are across the range of inputs they can encounter after deployment. 
This is particularly important as DRL agents make decisions based on Deep Neural Networks (DNNs) that function as black boxes with opaque input-output relationships from the perspective of human operators.
As such, we would like to reason about whether a DRL agent’s selected action $a$ changes in undesirable ways when the observed state $x$ is perturbed to $x + s$, where $s$ is a small, bounded slack.
Such properties are important, as the input to these agents typically comes from real-world measurements that are susceptible to noise caused by variability in measurement timing and limitations in measurement precision~\cite{dethise2021analyzing}.
Moreover, undesirable output deviations in response to small input changes can expose systems to adversarial manipulation, e.g., by inducing disproportionate resource allocations through minor input perturbations~\cite{liu2023exploring}.

\begin{figure}[t!]
    \setlength{\belowcaptionskip}{-8pt}
    \begin{center}
        \includegraphics[width=0.99\linewidth]{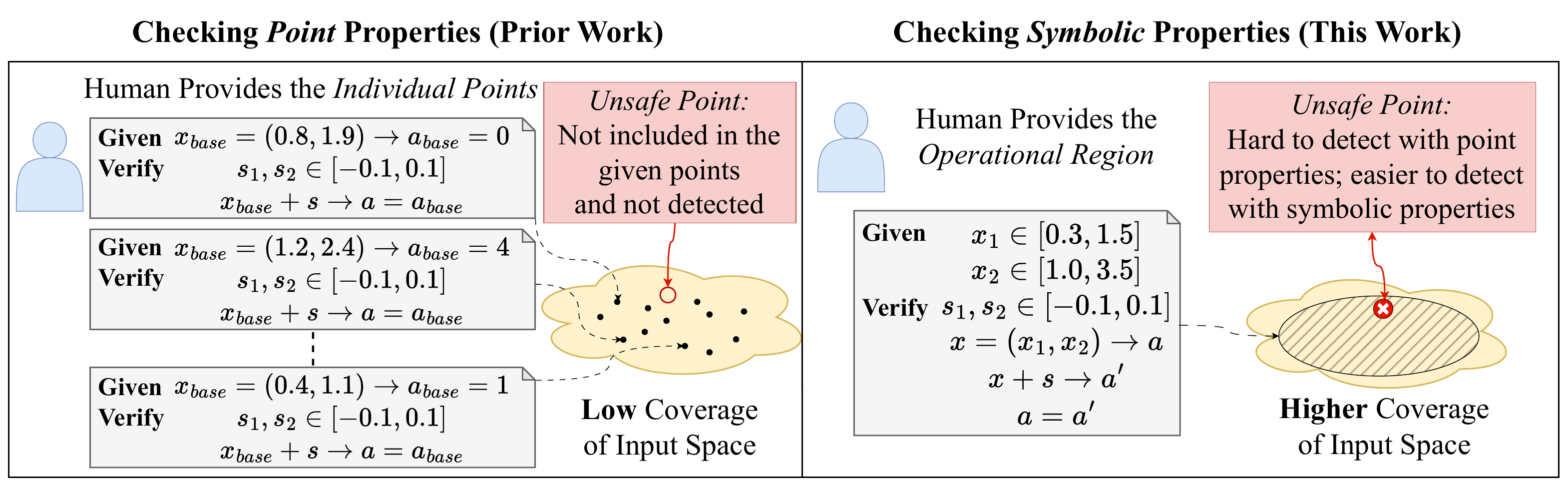}
        \caption{ \textbf{Point-wise versus symbolic property analysis.} 
        Left (Prior Work): Analysis is performed at individual points that humans provide, checking that the property holds under bounded perturbations around one concrete point in the input space. In this example, the property asserts that the output action does not change. Checking properties only at individual points results in limited coverage of the input space and may miss unsafe behavior outside the considered points.
        Right (This Work): Analysis is performed over an entire operational region using symbolic properties. This provides higher coverage of the input space and enables the detection of unsafe points that are difficult to uncover with point properties.
        }
        \Description{Comparison between point-based and symbolic property verification. Prior work verifies properties on individually provided input points, resulting in low coverage and potential missed unsafe behaviors. In contrast, this work verifies properties over an entire operational region, enabling higher coverage of the input space and improved detection of unsafe points.}
    \label{fig:symbolic}
    \end{center}
\end{figure}

Previous work~\cite{eliyahu2021verifying, dethise2021analyzing, huang2023toward} commonly relies on state-of-the-art DNN verification engines~\cite{marabou} or Mixed-Integer-Linear-Program (MIP) solvers~\cite{cplex} to reason about DRL agents in the systems and networking domain. However, these works can only check \emph{point properties}, where the system state $x$ and the reference action $a$ are \emph{fixed to concrete constants} (see Fig.~\ref{fig:symbolic}, left).
For example, consider Pensieve~\cite{mao2017neural}, a DRL agent for adaptive video streaming.
Its input includes the client's video buffer size and the measured throughput for the past streamed video segment. 
As its action, Pensieve chooses one of six possible bitrates from 300 Kbps to 4.3 Mbps for the next video segment.
The goal is to choose bitrates that enable smooth and high-quality video playback.
WhiRL~\cite{eliyahu2021verifying}, which uses the Marabou DNN verification engine~\cite{marabou}, and similar works~\cite{dethise2021analyzing, huang2023toward}, can only analyze Pensieve against point properties such as the following: If the video buffer has zero or one segment and the throughput for the past video segments is as low as a user-specified value (i.e., one concrete point among all possible system states), the DRL agent's output action should not be 4.3 Mbps (i.e., a concrete action).

While checking each individual point property is typically feasible and efficient, relying on point properties to reason about DRL agents operating in complex, dynamic systems and networks has two main drawbacks. 
First, point properties have low coverage. Each property only provides assurance around a single concrete point in the agent’s operational input space.
Second, one must explicitly identify which input points are worth checking.
This limitation is often manageable in supervised learning, where a (large) labeled data set explicitly captures the expected output value for a wide range of input points (e.g., AutoSpec~\cite{jin2024autospec}).
In DRL, however, there is \textit{no predefined ground truth for the expected behavior at a given state}, i.e., the best action is unknown prior to training and must be learned through interactions with the environment~\cite{sutton2018reinforcement}. 
As such, making point properties effective in this setting requires substantial domain expertise to identify meaningful points, which is difficult to scale to a level that provides sufficient coverage.

In this paper, we focus on \emph{symbolic properties} instead, where the system state $x$ and the reference action $a$ are symbolic (see Fig.~\ref{fig:symbolic}, right).
This shifts analysis from isolated input points to entire regions of the agent's operational space, substantially increasing coverage and removing the need for domain experts to manually enumerate a large number of representative input-output pairs.
The key enabling insight behind our approach is that many properties of interest for DRL agents can be expressed as a symbolic comparison between two executions of the same policy evaluated on interrelated symbolic states -- a base input and a perturbed one.

While conceptually simple, this comparison enables us to analyze symbolic properties using existing verification tools and, crucially, makes it practical to explore how far such analysis can be pushed for DRL agents in systems and networking.
Our idea is to systematically decompose symbolic properties into a finite set of sub-properties that compare concrete output neurons of the two agent copies.
Each sub-property remains symbolic over the input space but is more constrained than the original formulation.
As a result, existing verification techniques -- previously applied only to point properties -- can be directly leveraged for their analysis.
As our case studies demonstrate (\S\ref{sec:pensieve}-\S\ref{sec:aurora}), this decomposition is often sufficient to make symbolic analysis tractable in practice for representative DRL agents in the systems and networking domain.
The encoding of these sub-properties can deliberately be chosen to be generic and solver-agnostic so that it can apply to a variety of DRL agents in systems and networking and is compatible with multiple existing verification and analysis techniques, such as MIP~\cite{tjeng2017evaluating}, Satisfiability Modulo Theories (SMT)~\cite{katz2017reluplex, wu2024marabou}, Bound Propagation~\cite{zhang2018efficient, xu2020automatic}, and Branch-and-Bound (BaB)~\cite{bunel2020branch, xu2021fast, wang2021beta, kotha2023provably}. 
This flexibility allows us to analyze the same set of sub-properties across multiple verification engines and empirically assess their complementary strengths, enabling broader coverage than what would be possible with any single DNN verification technique alone.

Building on these insights, we conduct an empirical study of symbolic properties for DRL agents in systems and networking.
Specifically, we create a framework called \emph{\sys} that takes in a DRL agent model, the operational range of its input variables, and property details. It then uses the encoding and decomposition strategies proposed in this work to generate queries for backend verification engines.
We apply our approach across three representative control domains -- adaptive video streaming (Pensieve~\cite{mao2017neural}), wireless resource allocation (CMARS~\cite{zangooei2023flexible}), and congestion control (Aurora~\cite{jay2019deep}) -- covering both discrete and continuous action spaces.

Across these case studies, we analyze symbolic monotonicity and robustness properties over broad operational input ranges, examine how property satisfaction evolves during training, demonstrate the significance of the discovered counterexamples, study the impact of model size on verifiability, and compare the behavior of multiple verification backends, namely MIP-, SMT-, and BaB-based engines~\cite{gurobi, wu2024marabou, xu2020automatic, xu2021fast, kotha2023provably}.
Collectively, these results provide the first systematic view of how far symbolic property analysis can be pushed for integrating DRL agents in systems and networking, and which insights such analysis can -- and cannot -- reliably deliver in practice.

\clearpage

\isignpost{Summary of contributions} This paper makes the following contributions:
\begin{itemize}[leftmargin=20pt, topsep=2pt]
    \item We introduce a generic formulation of symbolic properties for DRL agents in systems and networking to capture expected high-level behaviors over ranges of system states (\S\ref{sec:prop}).
    \item We enable the analysis of symbolic properties with existing DNN verification engines via comparative encoding and decomposition into a finite set of sub-properties (\S\ref{sec:diffrl}).
    \item Using our property verification framework \sys, we conduct a systematic study of symbolic properties for DRL agents in adaptive video streaming, wireless resource allocation, and congestion control, examining verifiability, counterexamples, training dynamics, and the impact of model capacity and solver choice (\S\ref{sec:setup}-\S\ref{sec:aurora}).
\end{itemize}

\section{DRL Agents in Systems and Networking}
\label{sec:scope}

A DRL agent uses a DNN to represent its policy or value function, with parameters that specify how system states are mapped to actions or expected rewards.
During training, the agent observes the system’s state, takes an action, receives feedback in the form of a reward, and adjusts the DNN's parameters to maximize the rewards.
In systems and networking applications, the state is typically represented by a feature vector capturing performance indicators such as throughput, latency, packet loss, or resource utilization. The agent’s actions correspond to control decisions, such as allocating resources, scheduling jobs, tuning streaming video bitrate or congestion window sizes.

\isignpost{Deterministic policies} We focus on analyzing the DNN agent in its \emph{post-training form}, as the trained DNN is what defines the agent’s decision-making logic. This makes our approach applicable to a wide range of DRL agents, \emph{regardless of their specific training algorithm}. 
Moreover, we observe that DRL agents predominantly operate \emph{deterministically} in practical deployment scenarios in systems and networking.
Some agents are inherently deterministic at runtime, such as those trained using Q-learning~\cite{yan2021acc} or Deterministic Policy Gradient (DPG)~\cite{chinchali2018cellular, gui2024redte, qiu2020firm}.
Others are derived from stochastic policy-gradient methods~\cite{mao2016resource, mao2019learning, bojja2019learning, hu2023giph, jog2021one, mao2017neural, abbasloo2020classic, ma2022multi, dery2023queuepilot, tang2022abs, xu2023teal, park2024topfull, jay2019deep, ko2024edgeric}. 
That is, the DNN outputs a probability distribution over actions, representing the policy. During training, actions are sampled from this distribution to enable exploration, while during deployment, it is common practice to deterministically select the action with the highest probability~\cite{montenegro24learningDeterministicWithStochastic, eliyahu2021verifying}.
This deterministic behavior ensures reproducibility and predictability of the agent's decision-making behavior.

\isignpost{Common architectures and numerical action spaces} DNN architectures for DRL agents in systems and networking often consist of fully connected or convolutional layers with Rectified Linear Units ($ReLU(x) = \max(0, x)$) as activation functions. 
The action space is typically \emph{numerical}~\cite{mao2017neural, jia2023rdladder, jia2024dancing, wang2024autothrottle, zangooei2023flexible, qiu2020firm, yan2021acc, learning22wang, tang2022abs, dery2023queuepilot, chinchali2018cellular, jay2019deep, abbasloo2020classic, park2024topfull}, representing control decisions, such as the amount of resources to allocate to each component, the congestion window size, or the bitrate in video streaming applications.
This numerical action space can be discrete~\cite{mao2017neural, jia2023rdladder, jia2024dancing, wang2024autothrottle, zangooei2023flexible, qiu2020firm, yan2021acc, learning22wang, tang2022abs, dery2023queuepilot} or continuous~\cite{chinchali2018cellular, jay2019deep, abbasloo2020classic, park2024topfull}.
For agents operating in discrete action spaces (e.g., pick one of five video bitrates), the DNN typically provides one output neuron per possible action, each representing either the predicted Q-value (value-based methods) or the probability of selecting that action (policy-based methods). 
For continuous numerical action spaces (e.g., picking a memory size from a given range for resource allocation), the DNN output typically represents the parameters of a continuous probability distribution (e.g., the mean and variance of a Gaussian). Deterministic agents with continuous actions may also use a single output neuron per action dimension, directly representing the output action.
\clearpage

\section{Symbolic Properties}
\label{sec:prop}

To safely and effectively integrate DRL agents into control loops in systems and networking, it is essential to reason about how an agent’s selected action changes when the observed system state is perturbed. In practice, such perturbations may arise from noisy measurements or transient fluctuations in system conditions. A symbolic property captures whether these input perturbations cause the agent’s output to change in undesirable ways.

\subsection{Definition of Symbolic Properties for DRL Agents}

Formally, let $\pi: X \rightarrow A$ denote the deterministic policy function 
that determines the DRL agent's action, where $X \subseteq \mathbb{R}^n$ is the input state space, 
and $A\subseteq \mathbb{R}$ is the action space.
In our setting, $\pi$ is represented by a DNN. 
We define a symbolic property by specifying \emph{constraints over pairs of executions} of $\pi$. Specifically, for any input state $x \in X$, we compare the agent’s output at $x$ and at a perturbed state $x+s$, where a comparison function $f$ and a threshold $d$ capture the notion of acceptable change in the action.
Moreover, the allowable perturbations $s \in \mathbb{R}^n$ are restricted by per-dimension lower and upper bounds on elements of $s$:
{
\begin{equation}
\setlength{\abovedisplayskip}{5pt}
\setlength{\belowdisplayskip}{5pt}
\label{eq:general}
    \forall x \in X, l_{s_i} \leq s_i \leq u_{s_i}: \; \; f(\pi(x), \pi(x+s)) \leq d
\end{equation}
}

\begin{figure}[t!]
    \begin{center}
        \includegraphics[width=0.89\linewidth]{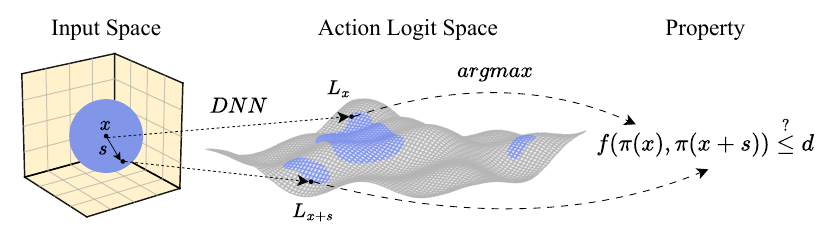}
        \caption{\textbf{Symbolic properties over DRL agent execution pairs (\S\ref{sec:prop}).} 
        A symbolic input $x$ from the region of interest in the input space and its bounded perturbation $x+s$ induce two executions of the agent's policy $\pi$. 
        The policy $\pi$ is based on a non-linear DNN. As such, input perturbations may result in jumps in the DNN's multi-dimensional continuous output space (see blue highlights in the center). The DNN's output may represent predicted action values or action probabilities, which we refer to as action logits in general. For deterministic agents with discrete actions (\S\ref{sec:scope}), the policy $\pi$ is defined as the $argmax$ of these logits. Our symbolic properties constrain the two resulting actions $\pi(x)$ and $\pi(x+s)$ with a function $f$ and tolerance $d$.}
        \Description{This figure illustrates the mapping from the input space to the action logit space through a deep neural network (DNN), and how properties are evaluated on the resulting policy. A small perturbation $s$ around an input $x$ is propagated through the DNN, producing corresponding logits $L_x$ and $L_{x+s}$. After applying the $\arg\max$ operator, the resulting actions $\pi(x)$ and $\pi(x+s)$ are compared via a property function $f(\cdot)$ to verify whether a specified constraint (e.g., bounded deviation $\leq d$) holds.}
    \label{fig:property}
    \end{center}
\end{figure}

In other words, the property is satisfied iff for all $x \in X$ and all perturbations $s$ within the specified bounds, the output comparison $f(\pi(x), \pi(x+s))$ is bounded by $d$.
Verifying properties of this form is not trivial, as the policy is based on a non-linear DNN and thus small changes in its input may lead to jumps in its output (see Fig.~\ref{fig:property}).

The comparison function $f$ and perturbation bounds are left abstract to allow the same symbolic formulation to capture a wide range of properties.
This allows properties to vary both in what aspects of the output are compared and how input perturbations are structured, without changing the underlying analysis pipeline. 
Based on common patterns in the choice of $f$ and the perturbation bounds, we instantiate this general formulation with two broad classes of symbolic properties that arise naturally in systems and networking: \emph{robustness} and \emph{monotonicity}.

\isignpost{Category 1: Symbolic robustness properties} We want to ensure that the DRL agent is \emph{robust}, meaning for any input in the agent's given operational range, small perturbations to the input should not cause disproportionately large changes in the output.
For example, if a DRL agent uses network packet loss rate as an input feature to decide the congestion window size, minor fluctuations in the measured loss rate (e.g., due to noisy measurements or small transient fluctuations in network conditions) should not result in substantial changes to the congestion window size.

Formally, our robustness property checks whether for any input $x\in X$, the agent's output changes by at most $d > 0$ when the perturbation has an $L_{\infty}$-norm bounded by a small $\epsilon > 0$.
This definition is consistent with prior work on observation robustness~\cite{moos2022robustRL}, which aims to find policies that are insensitive to small perturbations of their input observations. 
Within our property specification framework, this translates to setting all perturbation lower bounds to $-\epsilon$ and all upper bounds to $\epsilon$, and defining $f$ as the absolute value of the difference between its input actions: 
{
\setlength{\abovedisplayskip}{5pt}
\setlength{\belowdisplayskip}{5pt}
\begin{equation}
\label{eq:robustness}
    \forall i, l_{s_i} = -\epsilon, u_{s_i} = \epsilon \;\;\;\;\;\;\;\;\;\;\;\;\;\;
    f(y, z) = |y - z|
\end{equation}
}

\isignpost{Category 2: Symbolic monotonicity properties} Another important class of properties concerns \emph{monotonicity}, which captures expected directional relationships between input features (the system state) and the agent's output. Informally, monotonicity requires that, over the agent's given operational range, increasing a specific input feature should cause the output action to predictably increase or decrease in the expected direction. 
For instance, if the network loss rate increases, we expect a DRL-based congestion control agent to decrease the congestion window size.

Formally, we say that the policy $\pi$ is monotonically increasing with respect to its $i^{\text{th}}$ input feature if, for any input $x\in X$, increasing the $i^{\text{th}}$ component of $x$ by up to $\epsilon_i > 0$ results in an equal or greater output action with a tolerance of $d$. The tolerance parameter $d$ accounts for small valid output fluctuations and prevents the property from being overly restrictive. 
Within our property specification framework, this translates to (i) setting the perturbation lower bounds and all but the $i^{\text{th}}$ upper bounds to zero, (ii) setting $u_{s_i}$ to $\epsilon_i$, and (iii) defining $f$ as the directional change:
{
\setlength{\abovedisplayskip}{5pt}
\setlength{\belowdisplayskip}{5pt}
\begin{equation}
\label{eq:monotonicity}
    \forall j \neq i, l_{s_j} = u_{s_j} = 0 \;\;\;\;\;\;\;\;\;\;\;\;\;\;
    l_{s_i} = 0, u_{s_i} = \epsilon_i
    \;\;\;\;\;\;\;\;\;\;\;\;\;\;
    f(y, z) = z - y
\end{equation}}

The monotonically decreasing property is formulated analogously by flipping the direction of either $f$ or the bounds.
We provide several concrete instances of the robustness and monotonicity properties in the case study sections~\S\ref{sec:pensieve}-\S\ref{sec:aurora}.
Defining these properties does not require deep knowledge of the underlying DRL model or its training process. Instead, the properties encode expected relationships between system metrics and control actions that are already well understood by domain experts in systems and networking (e.g., higher loss should not increase sending rate). In practice, defining a monotonicity property amounts to selecting the relevant input feature, the expected direction of change, and appropriate tolerance parameters ($\epsilon_i$, $d$), while the logical structure of the property remains fixed and can be analyzed automatically in a symbolic manner.

\isignpost{Beyond robustness and monotonicity}
While our work focuses on robustness and monotonicity, as representative and widely applicable property classes, our formulation of symbolic properties is not limited to these instances and can express a variety of other properties through alternative choices of the comparison function $f$ and the perturbation bounds. 
The function $f$ does not necessarily need to be related to the difference between $\pi(x)$ and $\pi(x+s)$. 
It can be any linear combination of the two policy outputs or even compare different policies $\pi$ and $\pi^\prime$. 
For example, a policy can be compared with a safe baseline policy to find concrete input scenarios in which these two policies decide significantly different actions.
One can also leverage our general framework to explore richer properties by specifying directional perturbations for multiple input features. For example, when the network loss rate and measured latency increase at the same time, we expect a DRL-based congestion control agent to decrease the congestion window size.

Moreover, properties may also capture temporal trends when the input $x$ includes the history of a metric. For example, consider an agent that takes in the last $k$ bandwidth measurements ($x_1, \cdots, x_k$) to decide the data transmission rate.
Suppose we want to check a monotonicity property where decreasing bandwidth results in decreasing rate. However, we also want to make sure that if all $k$ measurements except for the most recent one have an upwards trend ($x_k < x_{k-1} < \cdots < x_2$), we do not consider the property violated if $x_1$ in $\pi(x+s)$ is lower than in $\pi(x)$.
In this case, we only need to adjust the input space $X$ by intersecting it with the negation of linear constraints that capture the relationship between the input variables in $x$, i.e., $x_k \ge x_{k-1} \vee x_{k-1} \ge x_{k-2} \vee \cdots \vee x_3 \ge x_2$.
Crucially, all such properties can be expressed as an instance of Eq.~\ref{eq:general} and handled by the same encoding and decomposition strategy described in \S\ref{sec:diffrl}.

\subsection{Further Considerations on Symbolic Properties}
\label{sec:property-considerations}

\isignpost{Accounting for multi-step behavior}
The symbolic properties defined above are based on a single-step execution $\pi(x)$ of a DRL agent. 
Because they are defined and verified over a bounded subset $X$ of the agent's input space, they capture the agent's behavior for any arbitrary hidden system state that results in inputs within the specified bounds.
Thus, the agent's decision for any such state, and also for any possible trajectory that may lead to such a state, is accounted for.

For some DRL agents, like Pensieve~\cite{mao2017neural} and Aurora~\cite{jay2019deep}, the input $x$ explicitly includes a history of metrics collected over the past $k$ steps.
Let these inputs be represented by $x_1,\, \ldots,\, x_k$.
The sequence of $x_i$ influences the chosen action, which in turn affects the subsequent state, and specifically, the next observed metrics in the history $x_j, j > i$.
As such, some histories are more likely to occur in practice, while others may be uncommon.
When it comes to safety properties, capturing more combinations (i.e., over-approximation) does not compromise safety guarantees. 
If the analysis verifies that an agent is safe, the agent will be safe even if the more infrequent combinations do not happen.
If the analysis returns a counterexample that is deemed uncommon or even represents desired behavior, it can be ruled out by further refining the input and slack bounds of the property.

\isignpost{Properties as safety checks, not effectiveness guarantees} While monotonicity and robustness properties are essential, satisfying them alone does not guarantee that a DRL agent is effective in its decision making. For instance, a DNN that outputs a constant action regardless of its inputs may trivially satisfy these properties, yet such a DRL agent is inherently ineffective and fails to achieve meaningful behavior. However, when these properties are applied to an agent that performs well in terms of the achieved rewards in the training scenarios, they can serve as critical safety checks. Specifically, they can help assess the generalizability and trustworthiness of a DRL agent beyond the specific scenarios encountered during training. For example, property violations often indicate deficiencies in the training procedure (such as overfitting) that can be addressed to improve the DRL agent. Consequently, while these properties do not guarantee effective decision making on their own, they are indispensable for developing DRL agents that are both effective and dependable.
We examine how property satisfaction evolves during training in our case studies (\S\ref{sec:pensieve}).

\section{Analysis of Symbolic Properties with \sys}
\label{sec:diffrl}

Once the user expresses a symbolic property in the form of Eq.~\ref{eq:general}, our goal is to determine whether it holds over the specified operational range of the system state variables $x\in X$.
Similar to prior work~\cite{eliyahu2021verifying, dethise2021analyzing, huang2023toward, jin2024autospec}, we seek to leverage existing DNN verification engines to perform this analysis. However, unlike prior approaches that verify properties for a fixed, concrete input $x$, our goal is to keep $x$ symbolic so that the result applies to all states within the specified range.

A practical challenge in doing so lies in representing the agent’s selected action $\pi(x)$ in a way that is compatible with existing verification tools.
As discussed in \S\ref{sec:scope}, many DRL agents in systems and networking~\cite{mao2017neural, jia2024dancing, yan2021acc, wang2024autothrottle, zangooei2023flexible, qiu2020firm, tang2022abs, dery2023queuepilot} operate over a finite, discrete numerical action space, where each output neuron corresponds to a valid control decision (e.g., the bitrate choice in Pensieve).
In practical deployment, these agents act deterministically -- either inherently or after collapsing stochastic policies -- by selecting the action corresponding to the maximum-valued output neuron, i.e., $\pi(x)$ is usually the $argmax$ over the output layer.

Existing DNN verification engines -- whether MIP-based~\cite{tjeng2017evaluating}, SMT-based~\cite{katz2017reluplex, wu2024marabou}, or BaB-based~\cite{bunel2020branch, xu2021fast, wang2021beta, kotha2023provably}-- 
are most naturally applicable to properties when the policy’s selected action is known a priori, i.e., when the property can be expressed by constraining a specific output of the network relative to the others. 
Intuitively, when the selected action is fixed, the property boils down to a relatively small set of linear inequalities between known outputs, which significantly constrains the search space. 
If the selected action is not fixed, the $argmax$ introduces a combinatorial, non-convex case distinction over all possible ``winning'' output logits, greatly increasing the number of regions the verifier must consider.
As a result, these verifiers do not natively support non-linear, discrete selection operators such as $argmax$ in a form amenable to symbolic reasoning.
This does not pose a problem for prior work on verifying DRL agents in systems and networking, as they instantiate $x$ to a concrete value and therefore fix which output neuron is selected by the policy.

In contrast, our symbolic properties quantify over the entire specified range of inputs. As a result, the selected output neuron, and consequently $\pi(x)$, depends on the input and cannot be fixed a priori. 
We address this mismatch by combining two ideas: (1) encoding symbolic properties by symbolically comparing two interrelated copies of the DRL agent, and (2) a simple decomposition strategy that reduces the symbolic property into a collection of sub-properties, each of which conditions on a fixed output action and can be analyzed using existing verification techniques.

This combination enables a practical advantage that is central to our case studies.
As we show in \S\ref{sec:pensieve}-\S\ref{sec:aurora}, existing verification engines exhibit complementary strengths across different sub-properties.
By keeping our encoding and decomposition strategy generic and solver-agnostic, we do not need to rely on a single engine to handle all cases -- different engines can resolve different sub-properties.
As we show in our case studies, this flexibility improves the practical tractability of analyzing symbolic properties and enables us to make progress beyond point properties for realistic DRL agents used in systems and networking.

\begin{figure}[t!]
    \setlength{\belowcaptionskip}{-16pt}
    \begin{center}
        \includegraphics[width=0.98\linewidth]{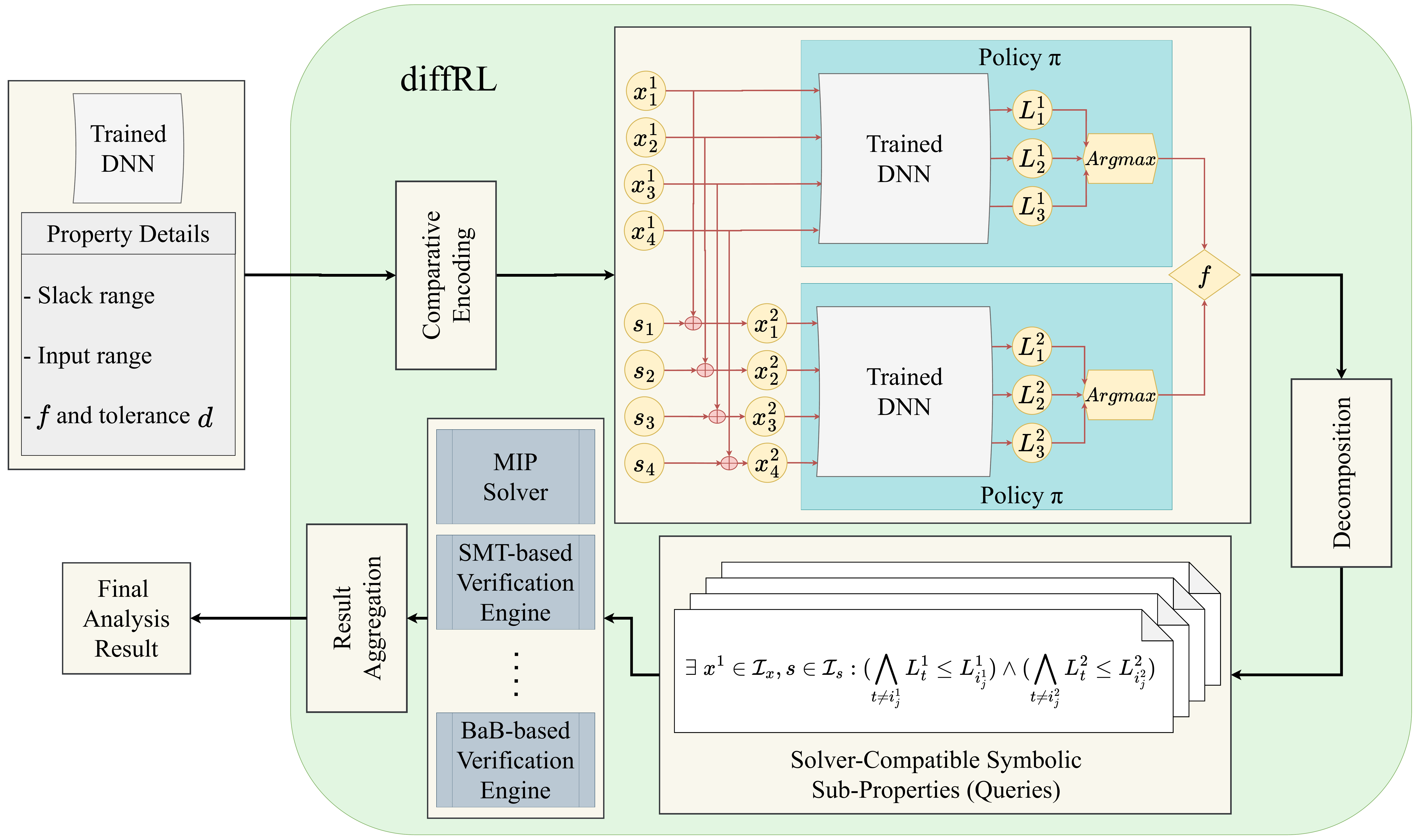}
        \caption{
        \textbf{\sys}'s overview (\S\ref{sec:diffrl}). Starting from a trained DNN and a symbolic property specification, \textbf{diffRL} applies a comparative encoding to construct two coupled executions of the same policy under a bounded input perturbation. The resulting formulation is decomposed into multiple queries, dispatched to heterogeneous solvers. The individual solver outcomes are then aggregated to produce the final verification result.
        }
        \Description{Overview of the diffRL framework: property specifications are encoded and decomposed into solver-compatible queries evaluated across multiple verification engines, with aggregated results yielding the final property assessment for the DNN policy.}
    \label{fig:design}
    \end{center}
\end{figure}

\isignpost{Comparative encoding} Given a property in the form of Eq.~\ref{eq:general}, we encode it by \emph{symbolically} comparing the output actions of two copies of the target DRL agent whose inputs are related through property constraints.
This is illustrated in Fig.~\ref{fig:design}.
Suppose the system state $x$ is a vector of size $n$. 
The input to the first copy, $x^1_1,\, \dots,\, x^1_n$ represents the \emph{base} scenario, where $x^1_i \in [l_{x_i}, u_{x_i}]$ is allowed to take any value within the operational range of the corresponding state variable, specified as lower and upper bounds $l_{x_i}$ and $u_{x_i}$.
The input to the second copy, $x^2 = x^1+s$, is the base scenario shifted by the slack vector $s$, whose elements are constrained under the property of interest. 
For example, the robustness property in Eq.~\ref{eq:robustness} constrains each element $s_i$ to be between $-\epsilon$ and $\epsilon$.

\isignpost{Property decomposition} Let $L^1$ and $L^2$ denote vectors of output neurons of the DNNs in the first and second copies, respectively (Fig.~\ref{fig:design}).
The agent's actions $\pi(x^1)$ and $\pi(x^2)$ correspond to the $argmax$ of $L^1$ and $L^2$, respectively.
Symbolic properties expressed using Eq.~\ref{eq:general} constrain the relationship between these two selected actions via the function $f$ and threshold $d$, i.e., they check whether
$f(\pi(x^1), \pi(x^2)) \leq d$ holds.
To enable analysis using DNN verification engines, we decompose this symbolic comparison into a finite set of sub-properties, each corresponding to a concrete pair of output neurons.

Specifically, for each pair of output neurons $(L^1_i, L^2_j)$, \sys automatically determines whether having $L^1_i$ and $L^2_j$ as the $argmax$ of the output neurons is a \valid or \invalid outcome. An \invalid outcome means that, given the assumed relationship between the two sets of inputs, selecting $L^1_i$ and $L^2_j$ as the $argmax$ of their respective DNN copies violates the property.
Similarly, a pair is a \valid outcome if this combination of selected actions is permitted by the property.
As a concrete example, consider a symbolic robustness property for Pensieve (see \S\ref{sec:pensieve}) that requires the output bitrate to change by no more than two resolution levels under small input perturbations.
If the output neuron corresponding to resolution $1440p$ is selected as the $argmax$ in the base copy, then output neurons corresponding to $480p$, $360p$, or $240p$ becoming the $argmax$ in the second copy would violate the property, and thus form invalid pairs with $1440p$.

Note that this decomposition is logically exact and does not introduce any approximation into the verification process. The symbolic property is violated \emph{if and only if at least one invalid pair is feasible} under the input and slack constraints.
For each invalid pair, \sys generates a query and invokes existing DNN verification engines to determine whether it is feasible.
If any invalid pair is feasible, \sys obtains a concrete counterexample for the symbolic property. Otherwise, the symbolic property is guaranteed to hold over the given operational input range.
Given a DRL agent and a property, \sys automatically generates the set of invalid output neuron pairs $\mathcal{P}=\{(i_1^1, i_1^2), \cdots, (i_m^1, i_m^2)\}$ for the two DNN copies.
For each invalid pair $(i_j^1, i_j^2)\in \mathcal{P}$, \sys creates a query $\mathcal{Q}_{j}$ that checks if any input to the DNN copies -- within the specified bounds -- leads to $i_j^1$ and $i_j^2$ being the $argmax$ in output layers $L^1$ and $L^2$, respectively:
{
\setlength{\abovedisplayskip}{4pt}
\setlength{\belowdisplayskip}{4pt}
\begin{equation} \label{eq:query}
    \mathcal{Q}_{j}: \exists \; x^1 \in \mathcal{I}_x, s \in \mathcal{I}_s:  (\bigwedge\limits_{t\neq i_j^1} L^1_t \leq L^1_{i_j^1} ) \wedge (\bigwedge\limits_{t\neq i_j^2} L^2_t \leq L^2_{i_j^2} )
\end{equation}
}
Here, $\mathcal{I}_x = [l_{x^1_1}, u_{x^1_1}]\times\cdots\times[l_{x^1_n},u_{x^1_n}]$ and $\mathcal{I}_s = [l_{s_1}, u_{s_1}]\times\cdots\times[l_{s_n},u_{s_n}]$ denote the bounds on the input state and slack variables.
If the property includes further constraints on the input space (see the temporal trends example in \S\ref{sec:property-considerations}), the constraints are added to decomposed properties in the queries as well. Each query is directly supported by existing DNN verification engines. If the verification engine finds a concrete set of input variables that satisfy the query constraints, \sys reports a violation of the symbolic property. If none of the queries for the \invalid pairs are satisfiable, the property holds and is reported as such by \sys.

\isignpost{Domain factors that make symbolic analysis tractable}
DRL agents used in systems and networking use DNNs that are substantially more compact than those used in other application domains~\cite{eliyahu2021verifying}.
Unlike domains such as computer vision, which require deep architectures to extract meaningful features from raw pixel data, systems and networking agents often operate on structured, high-level inputs like network latency, throughput, and buffer occupancy. As a result, their DNN architectures are shallower and contain fewer neurons.
This makes individual queries for these agents -- despite involving symbolic input ranges rather than single input points -- often tractable for existing DNN verification engines in practice.

Moreover, the size of the action space, which determines the number of output neurons, is usually small. For example, Pensieve~\cite{mao2017neural}, QueuePilot~\cite{dery2023queuepilot}, FIRM~\cite{qiu2020firm}, and CMARS~\cite{zangooei2023flexible} expose $6$, $6$, $15$, and $30$ actions, respectively.
This is not incidental: as the number of actions increases, DRL agents require substantially more data and computation to accurately estimate action values, limiting the model's ability to learn optimal policies in high-dimensional environments~\cite{sutton2018reinforcement}.
This is a manifestation of the curse of dimensionality~\cite{sutton2018reinforcement}. Consequently, practical DRL agents deliberately limit the action space.
As a result, the number of queries generated by decomposition remains manageable.

Together, these characteristics are the reason symbolic property analysis can be feasible for DRL agents in systems and networking.
By recognizing this and leveraging it through our comparative encoding and decomposition strategy, we are able to empirically explore how far symbolic analysis can be pushed for representative DRL agents in this domain (see \S\ref{sec:pensieve}-\S\ref{sec:aurora}).

\isignpost{Agents with continuous action spaces} 
Some DRL agents in systems and networking operate over continuous-valued action spaces. For these agents, the policy DNN does not select an action via an $argmax$ over discrete outputs. Instead, the DNN typically outputs the parameters of a continuous action distribution, from which the action is sampled. In practice, this distribution is most often chosen to be Gaussian, with its mean and variance produced by the DNN’s output layer.
In some implementations, the variance is held fixed and only the mean is learned~\cite{jay2019deep}. Concretely, letting $L_0$ and $L_1$ denote the output logits corresponding to the mean and variance, respectively, the action $\pi(x)$ is sampled from the normal distribution $\mathcal{N}(L_0, L_1)$. 

To support continuous action spaces, two adjustments to the definition and analysis of our symbolic properties are required.
First, since action selection does not involve an $argmax$ operation, there is no need for output decomposition.
Instead, our properties can be defined by comparing the distribution parameters produced by the DNN -- which the agents use to sample their actions -- and can be directly analyzed using existing DNN verification engines without decomposition.
This effectively allows robustness and monotonicity properties to be defined with respect to the expected action.
In other words, these properties compare the mean outputs of the two DNN copies under related inputs, in the same spirit as our comparative encoding for discrete-action agents.
Second, our symbolic properties can additionally include bounds on the distribution parameters of the first DNN copy to anchor the analysis to practically relevant regions in the action space.
One example of such property is $\exists \; x^1 \in \mathcal{I}_x,\; s \in \mathcal{I}_s,\; L^{1}_{0} \in \mathcal{I}_L :
    \big(|L^{1}_{0} - L^{2}_{0}| \le d \big)$, where $L^{1}_{0}$ and $L^{2}_{0}$ are the distribution means produced by the two respective DNN copies.
We use \sys to analyze properties for an agent with a continuous action space in \S\ref{sec:aurora}.

\section{Experimental Methodology and Solver Setup}
\label{sec:setup}

Leveraging \sys, we conduct a systematic study of symbolic robustness and monotonicity properties across three representative DRL agents that span different kinds of control loops in systems and networking: Pensieve~\cite{mao2017neural}~(\S\ref{sec:pensieve}), CMARS~\cite{zangooei2023flexible}~(\S\ref{sec:cmars}), and Aurora~\cite{jay2019deep}~(\S\ref{sec:aurora}).
This section describes the overall experimental setup for our case studies.

As described in \S\ref{sec:diffrl}, for each symbolic property, \sys generates a collection of verification queries via comparative encoding and decomposition.
Our case studies examine how these query sets behave in practice: which properties can be verified, where counterexamples arise, and how different verification backends perform across the resulting sub-properties, among other aspects.
Each query is analyzed with a timeout of \num{600} seconds and classified as \textsf{safe}, \textsf{unsafe}, or \textsf{unknown}. 
Recall that the queries generated from decomposition express \invalid cases that, if feasible, violate the property.
A query is considered \textsf{safe} if it is proven infeasible for all inputs in the specified range, \textsf{unsafe} if the verifier finds an input that makes it feasible, i.e., a counterexample that violates the symbolic property, and \textsf{unknown} if the verifier times out.

\isignpost{Backend Verification Engines} We use the following three backend DNN verification engines to analyze the queries generated by \sys (Eq.~\ref{eq:query}).
The engines have distinct strengths and limitations, and their performance can depend on preprocessing and configuration choices.

$\bullet\:$\emph{Marabou (SMT-Based).} Marabou~\cite{wu2024marabou} is an SMT-based DNN verification engine that can analyze properties specified as linear and piecewise-linear constraints. Its Deep Sum-of-Infeasibilities procedure handles piecewise-linear constraints via case analysis.
Marabou also supports a Split-and-Conquer (SnC) mode, which partitions a query into independent subproblems that can be analyzed in parallel. We use Marabou v2 in SnC mode with default settings.

$\bullet\:$\emph{Gurobi (MIP-based).} This approach encodes the DNN verification problem as a mixed-integer linear (MIP) that general-purpose solvers such as Gurobi~\cite{gurobi} or CPLEX~\cite{cplex} can analyze~\cite{tjeng2017evaluating}. 
The DNN’s affine transformations are modeled using real-valued variables and linear constraints, while binary variables encode the activation state of non-linear components such as $ReLU$.
To reduce the number of binary variables and improve verification time, we apply fast bound propagation techniques~\cite{xu2021fast} to identify $ReLU$ neurons that operate in fixed (active or inactive) regions prior to encoding each query. We use Gurobi under an academic license with 28 threads.

$\bullet\:$\emph{Alpha-Beta-CROWN (BaB-Based).} This approach combines bound propagation with systematic partitioning of the input or activation domains~\cite{bunel2020branch}. Bound propagation computes sound over-approximations of neuron values layer by layer under input constraints~\cite{zhang2018efficient, xu2020automatic} 
while branching recursively splits the domain into smaller subdomains.
Subdomains that are proven to satisfy the property are pruned; the remaining ones are further partitioned until either a counterexample is found or all subdomains are verified~\cite{bunel2020branch}.
We build on Alpha-Beta-CROWN~\cite{xu2020automatic, xu2021fast, kotha2023provably}, a state-of-the-art BaB-based verifier and VNN-COMP winner (2021--2023)~\cite{brix2023fourth}, which supports various branching and bounding strategies.
In our experiments, two features were particularly important for queries generated by \sys. Incorporating output constraints from our decomposition (Eq.~\ref{eq:query}) to tighten intermediate $ReLU$ relaxations substantially improves performance, building on recent advances using Lagrangian multipliers~\cite{kotha2023provably}. This is especially effective because our decomposed queries often impose multiple conjunctive constraints on the output layer. We also found that branching over the input domain is more effective than branching over $ReLU$ activation states. 
Tightening bounds using output constraints was proposed in~\cite{kotha2023provably} for $ReLU$-based splitting, and we extended Alpha-Beta-CROWN to support its combination with input-domain branching.

We selected these three verification backends because they represent the three primary algorithmic families in the modern DNN verification literature~\cite{muller2022third, brix2023fourth}. By choosing representative engines from each family, we aim to demonstrate the solver-agnostic nature of diffRL's encoding and investigate the complementary trade-offs between these approaches. For instance, while MIP solvers often excel on smaller, compact networks, BaB-based solvers like Alpha-Beta-CROWN leverage massively parallel bound propagation that can be more effective as model complexity scales. This selection allows us to evaluate how multi-engine verification improves overall tractability compared to relying on a single backend.
For each backend, we adopt solver-aware but conventional techniques -- such as query partitioning for Marabou, bound tightening before encoding for MIP, and combining input-domain splitting with output-constraint-based bound tightening for Alpha-Beta-CROWN -- to ensure that the queries produced by \sys are analyzed effectively.

All experiments were conducted on an Intel(R) Xeon(R) Silver 4314 machine with a 2.40GHz CPU. 
The trained models, property specifications, and verification engine configurations are publicly available at \url{https://github.com/mhmd97z/diffRL}.

\newcommand{\caseStudyPropertySkip}{\\[-0.5\baselineskip]}

\section{Case Study 1: Adaptive Video Streaming using Pensieve}
\label{sec:pensieve}

Adaptive bitrate (ABR) algorithms aim to improve user Quality of Experience (QoE) in video streaming by dynamically selecting the bitrate for each video segment, typically of length four seconds. Under varying network conditions, the objective is to maximize average bitrate while minimizing playback interruptions (rebuffering) and excessive bitrate fluctuations.

We investigate Pensieve~\cite{mao2017neural}, a widely studied DRL-based ABR agent that makes bitrate decisions based on a compact, structured representation of the system state.

\subsection{Agent Architecture}

The input features of the Pensieve agent comprise: (1) the bitrate selected for the previous video segment, (2) network throughput measurements for the past $k$ segments, (3) download times for the past $k$ segments, (4) the current playback buffer level, (5) the number of remaining video segments, and (6) the set of available bitrates out of six options for the next segment. We assume all bitrates are available.
Pensieve uses a history of the previous eight steps ($k = 8$) for two of these input features, resulting in a total of 25 input variables.
The policy network first computes separate embeddings for each of the six input features. These embeddings are then concatenated and passed through two fully connected (FC) layers with $ReLU$ activations, producing six output logits corresponding to the available bitrate options of $300, 750, 1200, 1850, 2850, \text{or } 4300$ kbps. Each $FC(H)$ layer contains $H$ output neurons, yielding the following architecture:
{\small
\begin{align*}
     \pi^{\text{Pensieve}}_{H}: \texttt{Input}(25) &\xrightarrow{{splitting}} [\texttt{Input}(1),  \texttt{Input}(8), \texttt{Input}(8), \texttt{Input}(1), \texttt{Input}(1), \texttt{Input}(6)] \\ & \rightarrow  [\texttt{FC}(H),\; \texttt{FC}(H),\; \texttt{FC}(H),\; \texttt{FC}(H),\; \texttt{FC}(H),\; \texttt{FC}(H)] \xrightarrow{{concatenation}} ReLU \\ 
    & \rightarrow \texttt{FC}(H) \rightarrow ReLU \rightarrow \texttt{FC}(6) \xrightarrow{argmax} \texttt{Output}(1)
\end{align*}
}

\subsection{Symbolic Properties}

Using our general framework from \S\ref{sec:prop}, we define the following symbolic properties for Pensieve.\caseStudyPropertySkip{}

\isignpost{Symbolic Property 1: Capacity Utilization (Monotonicity)}
If the available network throughput increases, an adaptive bitrate algorithm should not respond by selecting a lower video bitrate. We refer to this expected behavior as ``Capacity Utilization''.
We express this as a symbolic monotonicity property (see Eq.~\ref{eq:general} and Eq.~\ref{eq:monotonicity}) by constraining the slack vector so that only the input feature corresponding to measured throughput increases by at most $\epsilon$ while keeping all other input features unchanged. The property is violated if there exists an input state for which an increase in measured throughput causes the selected bitrate to decrease by more than $d$ levels.

\isignpost{Symbolic Property 2: Rebuffering Avoidance (Monotonicity)} When the playback buffer contains only a small number of video segments, the system is more vulnerable to transient throughput drops, making aggressive bitrate choices more likely to cause buffer depletion. In such situations, an adaptive bitrate algorithm should act conservatively.
We refer to this expected behavior as ``Rebuffering Avoidance''. Intuitively, for the same network conditions, the bitrate selected when the buffer is sparsely filled should not exceed the bitrate selected when the buffer is well filled.
We also capture this as a symbolic monotonicity property (see Eq.~\ref{eq:general} and Eq.~\ref{eq:monotonicity}) by constraining the slack vector so that only the input feature corresponding to buffer occupancy is allowed to increase and by at most $\epsilon$. The property is violated if there exists an input state for which increasing buffer occupancy leads to a decrease in the selected bitrate by more than $d$ levels.

\isignpost{Symbolic Property 3: Pensieve Robustness}
In practice, input features such as measured throughput and download time are subject to noise and small transient fluctuations. An adaptive bitrate algorithm should therefore avoid reacting to such minor perturbations with large changes in selected bitrate, as this can lead to unstable user experience.
We express this as a symbolic robustness property similar to Eq.~\ref{eq:robustness} by allowing small bounded perturbations (between $-\epsilon$ and $\epsilon$) across the input features. The property is violated if there exists an input state for which these perturbations cause the selected bitrate to change by more than $d$ levels.

\subsection{Analysis Results}

We use \sys to analyze the above symbolic properties for two configurations of Pensieve’s policy network, with hidden layer sizes $H=128$ and $H=64$. 
For all properties, we set the perturbation bound to $\epsilon=0.01$ and the tolerance to 
$d=3$ bitrate levels. With this choice of 
$d$, \sys generates 6 \invalid output neuron pairs -- and thus 6 verification queries\footnote{$(720,240),(1080,240),(1080,360),(1440,240),(1440,360),(1440,480)$} -- for each monotonicity property and $12$ for the robustness property, corroborating our insight that the number of generated queries from decomposition remains manageable in practice (\S\ref{sec:diffrl}).
Each invalid pair is analyzed independently using the backend DNN verification engines. Due to the agent's DNN architecture, not all backends are applicable to Pensieve. Specifically, the policy network computes separate embeddings for each input feature via input slicing, which is unsupported by Marabou. Consequently, we analyze Pensieve’s properties using the remaining two verification backends.

\begin{figure}[t!]
    \setlength{\belowcaptionskip}{-14pt}
    \setlength{\abovecaptionskip}{2pt}
    \begin{center}
        \includegraphics[width=0.999\linewidth]{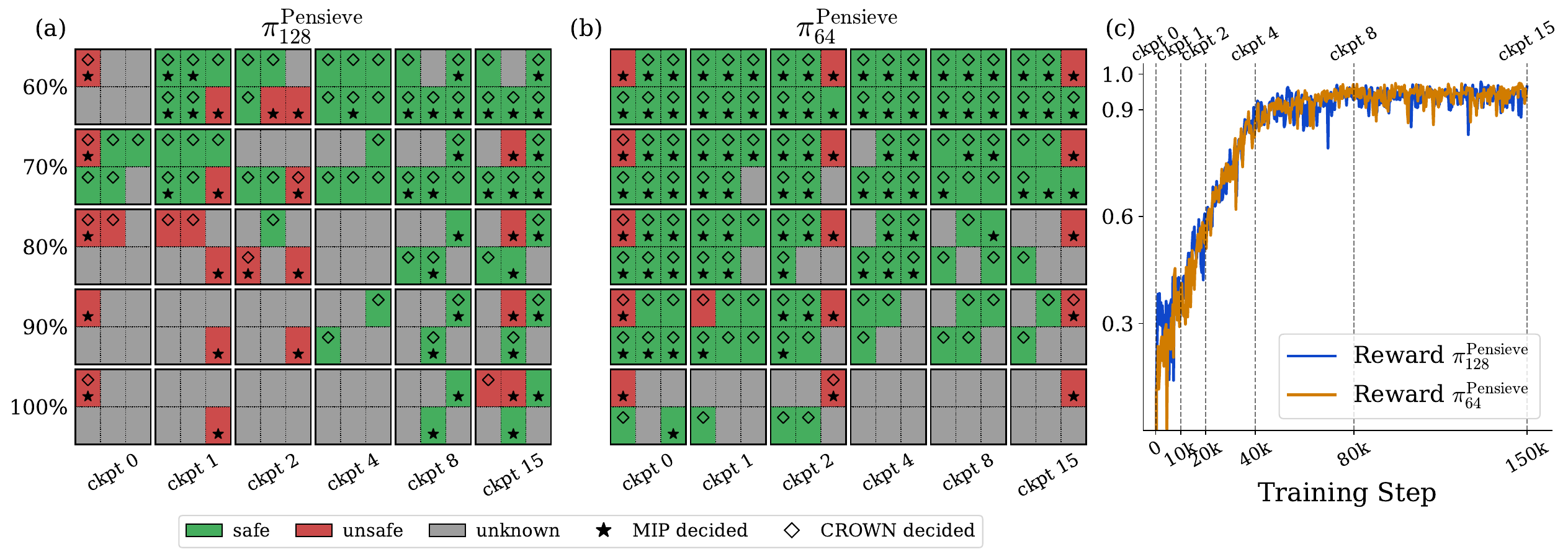}
        \caption{(a) and (b) illustrate the analysis results for the \emph{Capacity Utilization} property for DRL agents $\pi^{\text{Pensieve}}_{128}$ and $\pi^{\text{Pensieve}}_{64}$, respectively.
        Each group of six cells represents the results for a specific model checkpoint (\ckpt) throughout training (x-axis) and a specific per-input-feature coverage percentage (y-axis).
        Each cell in the group corresponds to one of the queries generated by \sys through property decomposition.
        (c) presents the training reward curves for $\pi^{\text{Pensieve}}_{128}$ and $\pi^{\text{Pensieve}}_{64}$, with vertical dashed lines indicating the checkpoints at which models are extracted for analysis.
        }
        \Description{This figure analyzes capacity utilization properties of Pensieve policies during training. Subfigures (a) and (b) show verification outcomes for models with 128 and 64 hidden units across checkpoints and input ranges (60\%-100\%), with results categorized as safe, unsafe, or unknown, and annotated by which solver (MIP or CROWN) provided a decision. Subfigure (c) plots the training reward over time for both models, illustrating the relationship between policy performance and property satisfaction across training.}
    \label{fig:pensieve_caputil}
    \end{center}
\end{figure}

Fig.~\ref{fig:pensieve_caputil} shows verification outcomes for the 6 queries generated for the Capacity Utilization property for the two Pensieve policies with different hidden-layer sizes of 128 ($\pi^{\text{Pensieve}}_{128}$) and 64 ($\pi^{\text{Pensieve}}_{64}$). 
In Fig.~\ref{fig:pensieve_caputil}(a) and (b), the x-axis corresponds to the model at a specific training checkpoint, where \ckpt~0 refers to the randomly initialized policy before any training and \ckpt~$i$ corresponds to the model checkpoint at $i\times10$k training steps.
The y-axis corresponds to the level of input-domain coverage, ranging from 60\% to 100\% of each feature's entire operational range\footnote{if an input feature's entire range is $[a,b]$, the queries corresponding to the $60\%$ level restrict this feature to the interval $[a+0.2(b-a), b-0.2(b-a)]$}.
For a certain checkpoint and coverage range, six individual cells are grouped together to represent the results for the property's six queries.
Cell colors indicate verification outcomes for individual decomposed queries (\textsf{safe}, \textsf{unsafe}, or \textsf{unknown}), where unsafe results correspond to concrete counterexamples in which higher measured throughput leads to a lower selected bitrate and \textsf{unknown} is reported when the verification engines time out before returning a conclusive result.
The symbols $\star$ and $\diamond$ indicate whether MIP and Alpha-Beta-CROWN successfully resolved the query, respectively.

\isignpost{Impact of input coverage} Across both model sizes, \sys successfully resolves most queries at coverage levels 60-80\%, producing \textsf{safe} and \textsf{unsafe} outcomes. As coverage increases toward 100\%, the fraction of \textsf{unknown} results grows, reflecting the increase in problem difficulty due to the larger input domain and expansion of the search space. Nevertheless, even in these challenging regimes, the analysis continues to yield conclusive results. 
Overall, these results demonstrate that symbolic property checking for DRL agents in systems and networking can be feasible and informative in practice, and can provide substantially broader coverage than individual point properties.

\isignpost{Training checkpoints reveal evolving property satisfaction}
For $\pi^{\text{Pensieve}}_{128}$, early training checkpoints exhibit a higher proportion of \textsf{unsafe} outcomes, indicating violations of the Capacity Utilization property in the initial stages of training. As training progresses, the fraction of \textsf{safe} outcomes increases, suggesting that the learned policy increasingly aligns with monotonic bitrate adaptation as throughput increases. Notably, a small number of violations persist even at the final checkpoints, and some queries that were previously verified as safe become unsafe after further training. 
This can be explained by the fact that continued updates to the model parameters during training change the policy’s decision behavior, which can invalidate earlier safety results. 

\begin{wrapfigure}{r}{0.5\textwidth}
    \setlength{\belowcaptionskip}{-14pt}
    \setlength{\abovecaptionskip}{2pt}
    \begin{center}
        \includegraphics[width=\linewidth]{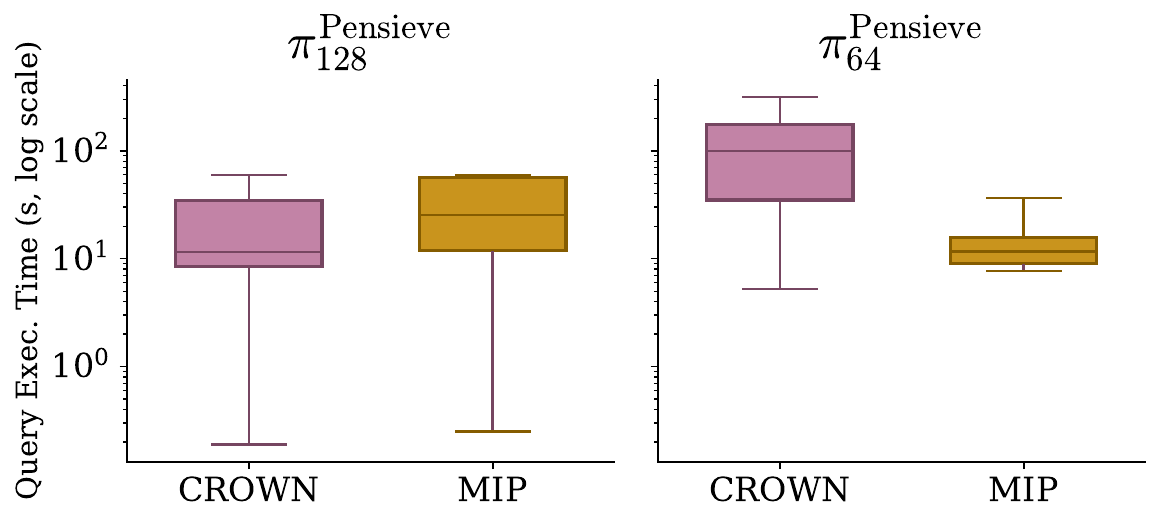}
        \caption{Query execution time comparison for verifying Capacity Utilization property of Pensieve policies with different model sizes and verification backends.
        }
        \Description{The figure presents the distribution of query execution times (in seconds, shown on a logarithmic scale) for four configurations: Pensieve-64 with CROWN, Pensieve-64 with MIP, Pensieve-128 with CROWN, and Pensieve-128 with MIP.}
    \label{fig:pensieve_caputil_exectime}
    \end{center}
\end{wrapfigure}

\isignpost{Impact of model size on verifiability and reward} 
Although $\pi^{\text{Pensieve}}_{128}$ with \num{103174} parameters and $\pi^{\text{Pensieve}}_{64}$ with \num{27142} parameters follow nearly identical reward trajectories (Fig.~\ref{fig:pensieve_caputil}(c)), their verification outcomes differ substantially. The smaller model produces approximately 45\% fewer \textsf{unknown} results, indicating that reduced model size can significantly improve symbolic verifiability without sacrificing the agent's effectiveness and performance. 
As such, this underscores the value of symbolic analysis as a complementary evaluation lens for DRL agents in systems and networking.

\isignpost{Benefits of multi-engine verification} Aggregating the outcomes across multiple verification engines substantially reduces the number of \textsf{unknown} results compared to relying on any single engine alone. For $\pi^{\text{Pensieve}}_{128}$, $\sim60\%$ of the resolved queries are decided by only one of the engines and the other timed out, whereas this fraction is about $35\%$ for $\pi^{\text{Pensieve}}_{64}$. These observations highlight the complementary strengths of different solvers and the necessity of multi-engine verification for enabling the analysis of symbolic properties, as relying on a single backend would leave many properties unresolved. 
This is especially true for larger models, whose verification search space is considerably more complex than that of smaller models. As shown in Fig.~\ref{fig:pensieve_caputil_exectime}, the substantial variability in query execution times across engines further underscores their complementary performance characteristics and the benefits of combining them.

\begin{figure}[t!]
    \setlength{\belowcaptionskip}{-12pt}
    \setlength{\abovecaptionskip}{3pt}
    \begin{center}
        \includegraphics[width=0.999\linewidth]{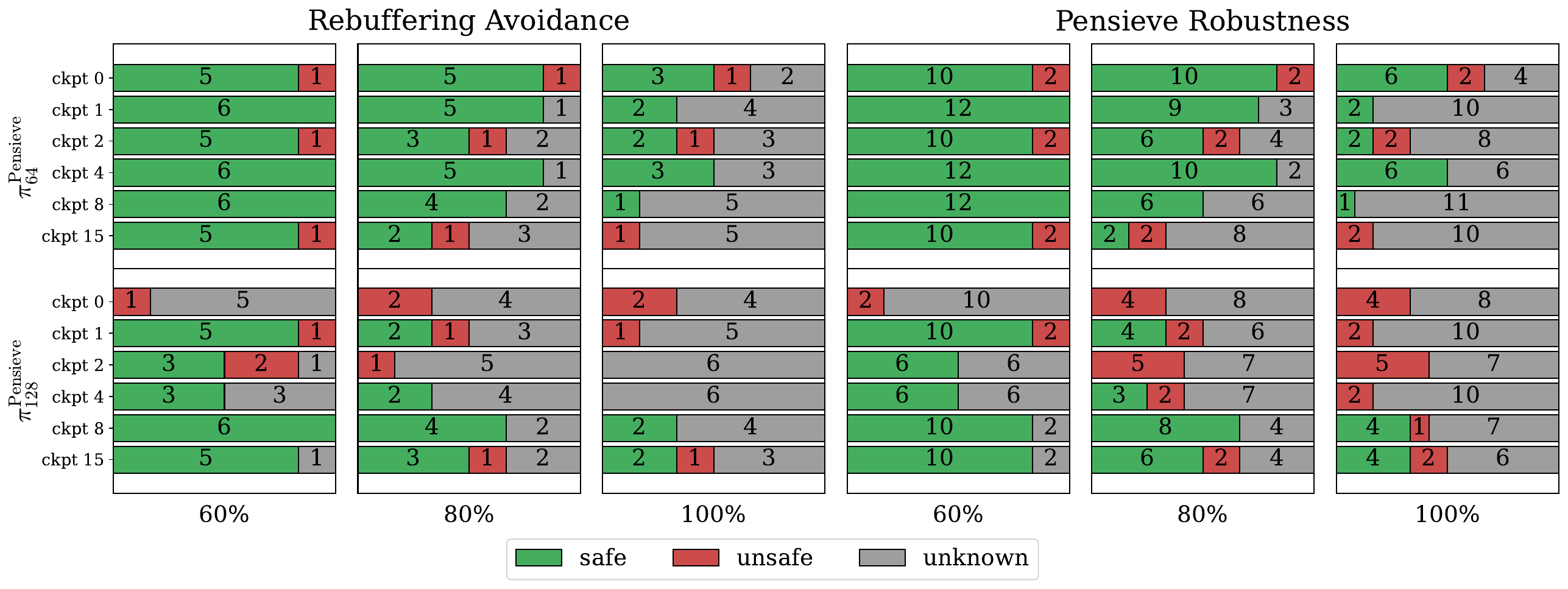}
        \caption{Analysis results for Rebuffering Avoidance (left) and Robustness (right) for $\pi^{\text{Pensieve}}_{64}$ (top) and $\pi^{\text{Pensieve}}_{128}$ (bottom) at different training checkpoints (rows) from Fig.~\ref{fig:pensieve_caputil}(c). Each column corresponds to a coverage level of the input domain (60\%, 80\%, and 100\%). For each model checkpoint and coverage, the stacked horizontal bars show the total number of verification queries classified as \textsf{safe} (green), \textsf{unsafe} (red), and \textsf{unknown} (gray), with the exact counts annotated inside each segment.}
        \Description{Verification results across training checkpoints for rebuffering avoidance and robustness properties in Pensieve models, showing the distribution of safe, unsafe, and unknown cases under varying input ranges.}
    \label{fig:pensieve_rob_bufavoid}
    \end{center}
\end{figure}

\isignpost{Other Pensieve properties} Fig.~\ref{fig:pensieve_rob_bufavoid} reports the analysis results for Rebuffering Avoidance and Robustness for $\pi^{\text{Pensieve}}_{64}$ and $\pi^{\text{Pensieve}}_{128}$ across training checkpoints and multiple input coverage levels. In contrast to Fig.~\ref{fig:pensieve_caputil}, which presents fine-grained, per-query outcomes, this figure aggregates the results at each checkpoint and coverage level, showing the total number of queries classified as \textsf{safe}, \textsf{unsafe}, and \textsf{unknown}.

Similar to Capacity Utilization, for both properties and both model sizes, most queries are conclusively resolved at 60\% and 80\% coverage, with a clear dominance of \textsf{safe} outcomes, especially for Rebuffering Avoidance. As the coverage increases to $100\%$, the fraction of \textsf{unknown} results grows substantially, particularly for the Robustness property. This mirrors the trend observed for Capacity Utilization, reflecting the rapid growth of the verification search space as the input domain expands. Nevertheless, even at full coverage, \sys continues to identify concrete \textsf{safe} and \textsf{unsafe} behaviors, highlighting the benefits of symbolic properties.

\section{Case Study 2: Wireless Resource Allocation using CMARS}
\label{sec:cmars}

Next-generation mobile networks allocate wireless resources across multiple network slices, where each slice groups users with similar service requirements and is governed by a service-level agreement (SLA). The network operator must allocate radio resources efficiently while ensuring that each slice meets the performance guarantees of its SLA.

We study CMARS~\cite{zangooei2023flexible}, a DRL agent that dynamically assigns radio resource blocks to individual slices with the objective of minimizing total resource usage while satisfying slice-level SLAs. 

\subsection{Agent Architecture}
The input features of the CMARS agent comprise: (1) recent SLA violation ratio, (2) current network quality, represented by the average signal-to-noise ratio (SNR) between users and the base station, computed per slice, (3) the amount of available radio resources, and (4) aggregated statistics from other slices, which include the number of Internet-of-Things users, the average traffic of constant-bitrate users, and the average traffic of variable-bitrate users. All input features are normalized to the range $[0,1]$ based on expected operational limits.
CMARS outputs a discrete action corresponding to the number of radio resource blocks allocated to the target slice, ranging from zero to the total number of available blocks $M$.

We analyze CMARS models with two architectural variants -- using either two or three fully connected layers -- and two action-space sizes, with $M \in \{15,30\}$ possible allocation levels. Each fully connected layer contains 32 neurons with ReLU activations, yielding the following architectures: 
{\small
\begin{align*}
    \pi^{\text{CMARS}}_{2, \; M}: \; & \texttt{Input(19)} \rightarrow \texttt{FC(32)} \rightarrow ReLU \rightarrow \texttt{FC(32)} \rightarrow ReLU \rightarrow \texttt{FC}(M) \xrightarrow{argmax} \texttt{Output}(1) \\
    \pi^{\text{CMARS}}_{3, \; M}: \; & \texttt{Input(19)} \! \rightarrow \! \texttt{FC(32)} \! \rightarrow \! ReLU \! \rightarrow \! \texttt{FC(32)} \! \rightarrow \! ReLU \! \rightarrow \! \texttt{FC(32)} \! \rightarrow \! ReLU \! \rightarrow \! \texttt{FC}(M) \xrightarrow{argmax} \texttt{Output}(1)
\end{align*}
}

\subsection{Symbolic Properties}

Using our general framework from \S\ref{sec:prop}, we define the following symbolic properties for CMARS.\caseStudyPropertySkip{}

\isignpost{Symbolic Property 1: Contention-Aware Allocation (Monotonicity)} 
In a sliced wireless network, increased resource demand from other slices places additional strain on the shared radio resources. As such, for a fixed target slice, CMARS should not increase its allocation when competing slices experience higher demand, and should ideally reduce it. We refer to this expected behavior as ``Contention-Aware Allocation''.
We encode this as a symbolic monotonicity property (Eq.~\ref{eq:monotonicity}), using a slack vector that permits only increases up to $\epsilon$ in the input features capturing aggregated demand from other slices, while the other features remain unchanged. The property is violated if there exists an input state in which increased cross-slice demand causes the second DNN copy to select an allocation that exceeds that of the first by more than $d$ resource units.

\isignpost{Symbolic Property 2: Channel Compensation (Monotonicity)} Poor channel conditions reduce the effective bitrate per radio resource block. To maintain slice-level SLAs under such conditions, a resource allocation policy should compensate by allocating additional radio resources to the affected slice.
We refer to this expected behavior as ``Channel Compensation''. We encode it as a symbolic monotonicity property (Eq.~\ref{eq:monotonicity}), where the slack vector only allows decreases up to $\epsilon$ in the input feature capturing channel quality (e.g., average SNR) while the rest of the features remain unchanged.
The property is violated if there exists an input state in which a degradation in channel quality causes the second DNN copy to select an allocation that is lower than that of the first by more than $d$ resource units.

\isignpost{Symbolic Property 3: CMARS Robustness} 
In practice, CMARS input features can be subject to measurement noise and small transient fluctuations. A well-behaved resource allocation policy should therefore avoid large changes in allocated resources in response to minor perturbations of its inputs, as such sensitivity can lead to unstable behavior and inefficient resource usage. Moreover, excessive sensitivity can expose the system to adversarial manipulation, where small input changes trigger disproportionate resource allocations.
We formalize this as a symbolic robustness property (Eq.~\ref{eq:robustness}), in which all input features are allowed to vary within a small bounded range ($\epsilon$), and the property is violated if there exists an input state for which these perturbations cause the second DNN copy to select an allocation that differs from that of the first by more than $d$ resource units.

\subsection{Analysis Results}

We use \sys to analyze the symbolic properties defined above across the two CMARS architectures $\pi^{\text{CMARS}}_{2, \; M}$ and $\pi^{\text{CMARS}}_{3, \; M}$ for $M = 15$ and $M = 30$.
For all properties, we set the perturbation bound to $\epsilon = 0.01$. We use tolerance values of $d = 8$ for $M = 15$ and $d = 16$ for $M = 30$ resource units, corresponding to moderate and large allocation changes relative to the action-space size. 
With these parameters, \sys generates $28$ and $105$ \invalid output neuron pairs for the two monotonicity properties when $M = 15$ and $M = 30$, respectively; the number of queries is doubled for the robustness property. This again highlights that, while the number of queries grows with the action space, decomposition remains tractable in practice.

\begin{figure}[t!]
    \setlength{\abovecaptionskip}{2pt}
        \includegraphics[width=0.99\linewidth]{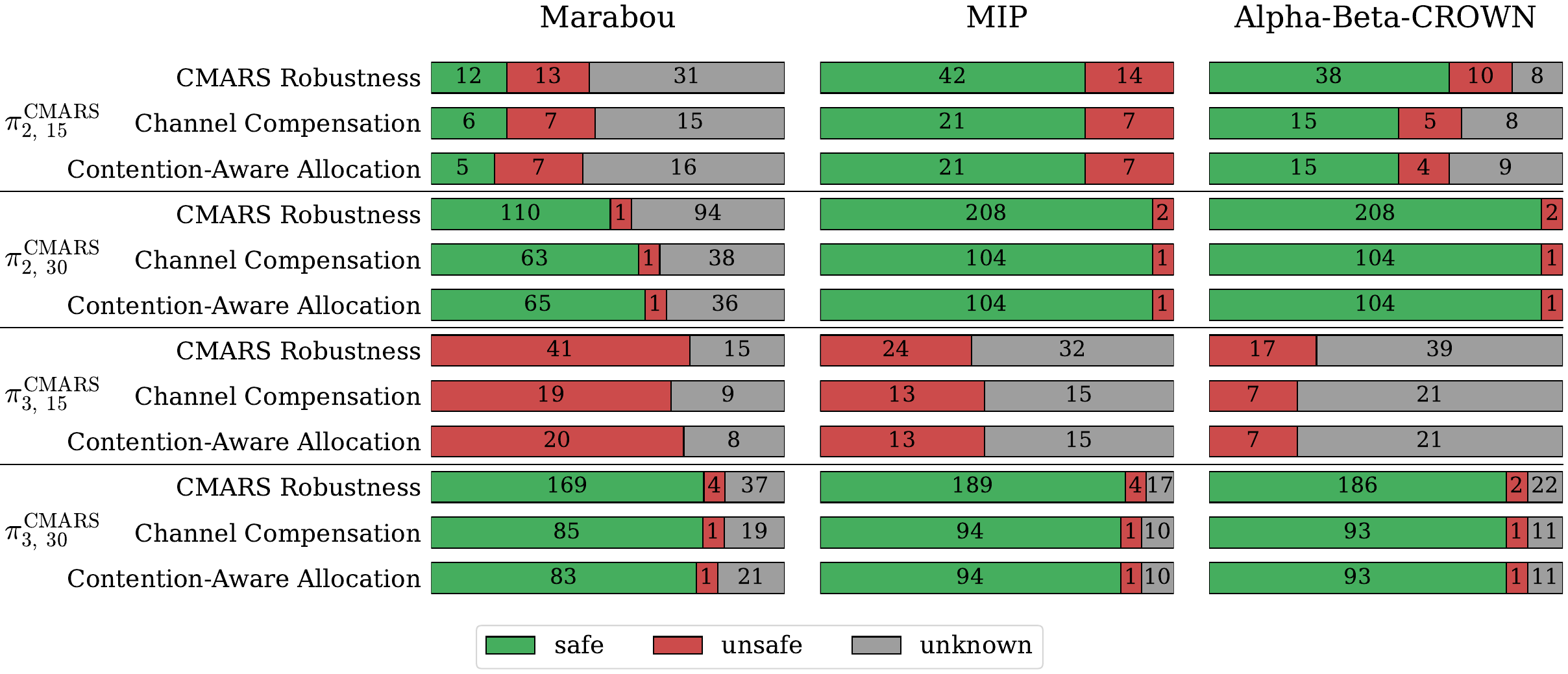}
        \caption{Comparison of analysis outcomes across solvers and properties. The figure reports the number of \textsf{safe}, \textsf{unsafe}, and \textsf{unknown} results obtained using Marabou, MIP, and Alpha-Beta-CROWN for queries generated by \sys for three properties evaluated on different CMARS policies. Each horizontal bar corresponds to a property-policy pair, with segment lengths and overlaid counts indicating the solver's outcome distribution.}
        \Description{Comparison of verification outcomes (safe, unsafe, unknown) across multiple solvers for CMARS policies under different configurations and property types.}
    \label{fig:cmars_results}
\end{figure}

Fig.~\ref{fig:cmars_results} summarizes the analysis for CMARS across different properties, policy configurations, and verification backends.
Each horizontal bar corresponds to a property-policy pair, with colored segments showing how many queries each solver classifies as \textsf{safe}, \textsf{unsafe}, or \textsf{unknown}.

\isignpost{Significance of counterexamples}
For both \emph{Contention-Aware Allocation} and \emph{Channel Compensation}, the identified counterexamples correspond to corner-case but operationally meaningful scenarios that appear to be underrepresented in the training data. These cases highlight potential weaknesses in the model's generalization that are unlikely to surface through standard evaluation metrics. 
The robustness counterexamples are particularly striking. 
In one representative counterexample in $\pi^{\text{CMARS}}_{2, \; 30}$, $\epsilon = 0.001$ induces a shift of $26$ resource units, exceeding $80\%$ of the available units in the $30$-action model. The corresponding input state is characterized by low-valued features, reflecting a lightly loaded scenario with minimal traffic demand, low resource utilization, no prior SLA violations, and only a small number of active devices. Under this state, the nominal policy selects $3$ resource units, whereas the perturbed input leads to the selection of $29$ units.

We also observe robustness violations in other operating regimes. In $\pi^{\text{CMARS}}_{2, \; 15}$, a counterexample arises where queues are moderately occupied (approximately half-full), a significant number of devices are active, resources are partially allocated, and recent SLA violations are non-negligible. In this case, the base input scenario yields an $argmax$ at $9$ resource units, with a corresponding logit of $0.705$. However, under a small admissible perturbation, the selected action shifts to $1$ resource unit, corresponding to an action distance of $8$. Notably, the logit of action $9$ under the perturbed input drops to $-0.018$, indicating a substantial change in the model’s preference ordering.

Such a disproportionate response to a minor input change indicates a high sensitivity to small fluctuations and highlights potential vulnerability to measurement noise or adversarial manipulation.
These counterexamples illustrate the practical value of symbolic analysis: they expose rare but impactful behaviors that are difficult to uncover through point-based analysis. We envision leveraging such counterexamples to guide targeted retraining and robustness-aware learning, as explored in recent work on counterexample-guided DRL refinement~\cite{boetius2024counterexample, gangopadhyay2023counterexample} (see \S\ref{sec:future}).

\isignpost{Deeper networks are not necessarily safer}
Policies $\pi^{\text{CMARS}}_{2,30}$ and $\pi^{\text{CMARS}}_{3,30}$ exhibit the highest levels of property compliance across the evaluated criteria, with relatively few violations and a large fraction of \textsf{safe} outcomes, suggesting that these models mostly behave reliably with respect to the analyzed properties.
In contrast, $\pi^{\text{CMARS}}_{3,15}$ shows substantially more violations and a higher number of \textsf{unknown} results, particularly for the \emph{CMARS Robustness} property. Notably, this model performs considerably worse than its ``shallower'' counterpart $\pi^{\text{CMARS}}_{2,15}$.
These results indicate that increasing network depth does not necessarily improve symbolic property compliance. Moreover, deeper models tend to be more computationally challenging to analyze, as reflected by the larger fraction of \textsf{unknown} outcomes, echoing a similar trend observed in the Pensieve case study.

\begin{figure}[h]
    \setlength{\belowcaptionskip}{-14pt}
    \setlength{\abovecaptionskip}{2pt}
    \begin{center}
        \includegraphics[width=0.99\linewidth]{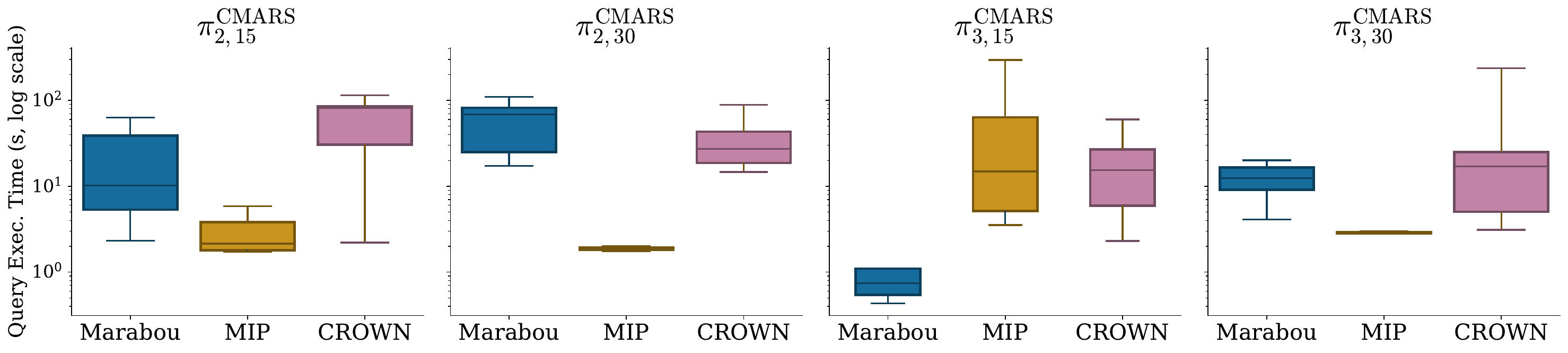}
        \caption{Query execution time (in seconds, log scale) for verifying the channel compensation property of CMARS policies under different architectures and action-space sizes using three verification backends (Marabou, MIP, and Alpha-Beta-CROWN). Each subplot corresponds to a policy.}
        \Description{The figure presents four boxplots comparing verification runtimes across solvers for CMARS policies with varying depth and action-space size, evaluated on the channel compensation property. The y-axis shows execution time on a logarithmic scale, while the x-axis lists the verification backends.}
    \label{fig:cmars_channel_exectime}
    \end{center}
\end{figure}

\isignpost{Benefits of multi-engine verification}
For smaller CMARS models ($\pi^{\text{CMARS}}_{2, \; 15}$ and $\pi^{\text{CMARS}}_{2, \; 30}$), the MIP-based approach is the only one that resolves all queries without timeouts, demonstrating its reliability for compact DNNs in this case study.
As model complexity increases, for $\pi^{\text{CMARS}}_{3, \; 15}$ and $\pi^{\text{CMARS}}_{3, \; 30}$, MIP begins to encounter scalability limitations and times out on a subset of queries. 
In this regime, different engines expose complementary strengths.
For $\pi^{\text{CMARS}}_{3, \; 15}$, Marabou identifies up to 30\% more violations compared to MIP, while MIP performs better than Marabou for the more compact $\pi^{\text{CMARS}}_{2, \; 15}$ counterpart with over $1000$ fewer parameters.
Conversely, for $\pi^{\text{CMARS}}_{3, \; 30}$, MIP proves more queries to be \textsf{safe} than Marabou while detecting the same number of violations.
Alpha-Beta-CROWN terminates on all queries for $\pi^{\text{CMARS}}_{3, \; 30}$, which meets the properties in most cases.
While it verifies a larger fraction of queries as \textsf{safe}, it detects fewer violations than Marabou. This trend is consistent across most CMARS configurations. Also, the differences across backends are reflected in query execution times for different policies (see Fig.~\ref{fig:cmars_channel_exectime}).

Overall, these results reinforce the benefits of a general symbolic property formulation and decomposition strategy that is compatible with multiple verification engines.
Different backends excel at different aspects of the verification task, and relying on a single engine would leave a non-trivial fraction of queries unresolved.

\section{Case Study 3: Congestion Control using Aurora} \label{sec:aurora}

Congestion control algorithms regulate the sending rate of an end-to-end transmission depending on the observed network conditions and anticipated trends. Their goal is to make efficient use of the available resources, i.e., achieving high data rates, low delays, and low packet loss rates.
We investigate Aurora~\cite{jay2019deep}, a DRL-based congestion control agent that operates on a continuous action space, making it qualitatively different from the discrete-action agents studied earlier.

\subsection{Agent Architecture}
The Aurora agent observes recent traffic statistics and continuously adjusts the sender’s rate in response.
At each decision step, Aurora’s policy network takes as input a vector of recent measurements capturing: (1) the latency ratio (current latency normalized by the minimum observed latency), (2) the packet acknowledgment ratio by destination (packets acknowledged normalized by packets sent), and (3) the latency gradient, indicating whether latency is increasing or decreasing.
These signals are collected over the previous $k$ decision steps, yielding an input of size $3k$.
The original work shows that a model with $k=2$ performs similarly to one with $k=10$. 
We select $k=3$ to keep the size of the input low.

At each decision step $t$, Aurora's DNN outputs the mean of a normal distribution, from which a continuous action $a_t$ is sampled.
The standard deviation is fixed to a constant, $\sigma = 0.5$ in this case. The sign of the action $a_t$ determines the direction of rate adjustment (increase or decrease) and its magnitude controls the adjustment extent. Specifically, if the previous sending rate is $x_{t-1}$, the updated rate $x_t$ is set to $x_{t-1} \cdot (1 + \alpha a_t)$ for $a_t \geq 0$, and to $x_{t-1} / (1 - \alpha a_t)$ otherwise.
The original Aurora implementation uses hyperbolic tangent activations, which are incompatible with many DNN verification engines. Following prior verification work~\cite{eliyahu2021verifying}, we replace these with ReLU activations in the following architecture, and use a retrained model from~\cite{qiu2024flash} that obtains comparable performance:
{\small
\begin{align*}
    \pi^{\text{Aurora}}_{128}:  \texttt{Input(9)} &  \rightarrow \texttt{FC}(128) \rightarrow ReLU \rightarrow \texttt{FC}(128) \rightarrow ReLU \rightarrow \texttt{FC}(128) \rightarrow \texttt{Output}(1)
\end{align*}
}

\subsection{Symbolic Properties}

Using our general framework from \S\ref{sec:prop}, we define the following symbolic properties for Aurora.\caseStudyPropertySkip{}

\isignpost{Property 1: Ack-Driven Capacity Utilization (Monotonicity)}
An increasing packet acknowledgment ratio indicates that a larger fraction of transmitted packets is successfully delivered, reflecting favorable network conditions. Under such conditions, Aurora should not reduce its sending rate, and should ideally increase it. We refer to this expected behavior as ``Ack-Driven Capacity Utilization''.
We encode this as a symbolic monotonicity property by allowing a positive slack (at most $\epsilon$) only on the packet acknowledgment ratio input features and checking if the resulting rate adjustment \emph{reverses direction} from increasing to decreasing.

Because Aurora selects actions by sampling from a distribution whose mean is produced by the DNN as described in \S\ref{sec:diffrl}, we define the property over the DNN outputs that determine these means (the standard deviation is fixed to $\sigma$). Concretely, we flag a potential violation when the means of the two DNN copies are separated such that the sign of the sampled action would differ with non-negligible probability across executions.
Formally, following the notation in \S\ref{sec:prop} and \S\ref{sec:diffrl}, we set $f(x, y) = x - y$, where $x$ and $y$ will be $L^1_0$ and $L^2_0$, the respective output means of the two DNN copies.
The threshold is set to $d = 2 \times \mu$, and the bound for $L^1_0$ to $[\mu, \infty)$.
This effectively means that ``invalid'' outputs occur when $L^1_0 \geq \mu \wedge L^2_0 \leq -\mu$.
Under standard distributional assumptions, if the above inequalities hold, the probability of the selected action's sign changing from positive in the first DNN copy to negative in the second is $\Phi (\frac{\mu}{\sigma})^2$, where $\Phi$ is the standard normal CDF (\S\ref{app:probab}). 
In our experiments, we set it to half the standard deviation $\frac{\sigma}{2}$, resulting in a non-negligible probability of about $40\%$ for a change in action direction.

\isignpost{Property 2: Latency-Aware Capacity Utilization (Monotonicity)} 
A lower latency ratio indicates that the current end-to-end latency is close to the minimum observed latency, suggesting lighter-filled queues along the path. 
As such, when the latency ratio decreases, Aurora should not reduce its sending rate, and should ideally increase it to better utilize available bandwidth. We refer to this expected behavior as ``Latency-Aware Capacity Utilization''.
We encode this similarly to the previous property, but with negative slack for the latency ratio inputs and check if the rate adjustment would change direction from increasing to decreasing.

\isignpost{Property 3: Aurora Robustness} 
This property captures the expectation that small perturbations in Aurora's observed state -- arising from noise, measurement variability, or transient fluctuations -- should not cause qualitatively different control decisions. 
Specifically, we check if bounded perturbations (up to $\epsilon$) \textit{across all input features}, including historical observations, can cause a reversal in the direction of the rate adjustment.
Because robustness concerns both possible direction changes, we consider two symmetric cases: a change from increasing to decreasing, and the converse. Each case is analyzed separately using the same comparative encoding as in the monotonicity properties.

\begin{table*}[t]
\small
\centering
\caption{Aurora analysis results for different solvers and per-input-feature coverage ranges.}
\label{table:aurora_results}
\begin{tabularx}{\textwidth}{
    p{4.6cm}
    >{\centering\arraybackslash}m{1.5cm}
    >{\centering\arraybackslash}m{1.5cm}
    >{\centering\arraybackslash}m{1.5cm}
    >{\centering\arraybackslash}m{3.5cm}
}
\toprule
Property & Coverage & Marabou & MIP & Alpha-Beta-CROWN \\
\midrule

\multirow{2}{*}{Aurora Robustness}
  & 70\%  & {\color{darkgreen}\textsf{safe}}   & {\color{darkgreen}\textsf{safe}}    & {\color{darkgreen}\textsf{safe}} \\
  & 100\% & \textsf{unknown} & {\color{red}\textsf{unsafe}}  & \textsf{unknown} \\
\addlinespace[1ex] \hline \addlinespace[1ex]

\multirow{2}{*}{Ack-Driven Capacity Utilization}
  & 70\%  & {\color{darkgreen}\textsf{safe}}   & {\color{darkgreen}\textsf{safe}}    & {\color{darkgreen}\textsf{safe}} \\
  & 100\% & \textsf{unknown} & {\color{red}\textsf{unsafe}}  & \textsf{unknown} \\
\addlinespace[1ex] \hline \addlinespace[1ex]

\multirow{2}{*}{Latency-Aware Capacity Utilization}
  & 70\%  & {\color{darkgreen}\textsf{safe}}   & {\color{darkgreen}\textsf{safe}}    & {\color{darkgreen}\textsf{safe}} \\
  & 100\% & \textsf{unknown} & {\color{red}\textsf{unsafe}}  & \textsf{unknown} \\
\addlinespace[1ex]
\bottomrule
\end{tabularx}
\end{table*}

\subsection{Analysis Results} 
Table~\ref{table:aurora_results} summarizes the analysis outcomes for Aurora’s symbolic properties under a perturbation bound of $\epsilon = 0.01$, considering $70\%$ and $100\%$ coverage of each input feature.
Similar to Pensieve~(\S\ref{sec:pensieve}), we observe that all engines can resolve the queries for $70\%$ per-input-feature coverage, with all three properties verified as \textsf{safe}.
As coverage increases to 100\%, the MIP-based approach is the only method that resolves all queries before the timeout period, identifying concrete \textsf{unsafe} behaviors for all three properties.

Besides demonstrating how our approach can extend to continuous action spaces, these results reinforce two broader insights. First, analysis coverage over the input space plays a critical role in revealing problematic behaviors. Second, with the current state-of-the-art verification techniques, relying on a single engine is not sufficient for analyzing symbolic properties.
A generic symbolic property formulation that is compatible with multiple solvers enables users to push analysis of such properties over as wide a range as possible.

\section{Discussion and Future Work} 
\label{sec:future}

\isignpost{Verifiability and Scalability}
The tractability of symbolic analysis is strongly influenced by both the structural properties of the model and the scope of the verification task. From a model perspective, smaller and shallower architectures are consistently easier to verify. For instance, in Pensieve, reducing the hidden dimension from $128$ to $64$ yields up to $45\%$ fewer unknown (timeout) outcomes while preserving comparable reward performance. Furthermore, larger or deeper architectures, such as those used in CMARS, introduce additional non-linearities and activation regions, increasing solver burden without necessarily improving compliance with symbolic properties.

Beyond model size, the scope of symbolic analysis, particularly the size of the input domain, plays a critical role in scalability. As the coverage of the input space expands (e.g., from $60\%$ to $100\%$ per dimension), the number of unknown results grows significantly due to the exponential increase in the solver’s search space. Importantly, our results show that verifiability is not solely determined by architecture or input bounds, but also by the specific function represented by the trained DNN. For a fixed model and fixed input coverage, the ratio of unknown queries varies substantially across training checkpoints. Early checkpoints often exhibit higher violation rates, whereas later checkpoints, despite aligning better with expected domain behaviors such as monotonicity, can lead to more complex decision boundaries that are harder for solvers to resolve. This variation in unknown ratios across checkpoints demonstrates that the geometry of the learned function itself directly impacts solver tractability. Overall, verifiability depends on the interplay between model architecture, input-domain specification, and the evolving complexity of the learned policy.

\isignpost{Multi-Engine Verification Performance} The results suggest that no single verification engine is superior across all scenarios, and a solver-agnostic approach is essential. Aggregating results from multiple backends (MIP, SMT-based Marabou, and BaB-based Alpha-Beta-CROWN) substantially reduces unknown outcomes by leveraging their complementary strengths. In particular, MIP-based solvers are effective on tightly constrained queries with limited activation ambiguity, and BaB-based methods excel when strong bound propagation and input-domain partitioning can prune large portions of the search space.

\isignpost{Automatic property identification} In this work, all properties are defined based on expert knowledge. An interesting avenue for future research is the automatic identification of properties across diverse domains. One potential approach for monotonicity properties would be to analyze empirical gradients over state-action distributions sampled from real-world traces. Features that exhibit consistently positive or negative gradients across observed data would indicate a strong directional influence on the agent's decisions and could be automatically flagged as candidates for formal symbolic monotonicity checks. This would reduce the reliance on manual domain expertise and allow for a more systematic discovery of operationally meaningful properties to verify.

\isignpost{Broader applicability to DRL agents in systems and networking}
While our evaluation focuses on three representative DRL agents, the symbolic property formulation enabled by \sys applies more broadly to DRL agents in systems and networking.
We briefly illustrate how similar monotonicity and robustness properties naturally arise in other settings.

Consider QueuePilot~\cite{dery2023queuepilot}, which is a DRL-based Active Queue Management agent aiming to address the challenge of managing small buffers in backbone routers. It controls the probability of Explicit Congestion Notification (ECN) marking to balance link utilization, packet loss, and queueing delay. Its input features capture traffic intensity, link utilization, proportion of marked packets, and queue occupancy, delay, and loss statistics.
Its discrete action space consists of a set of ECN marking probabilities.
In this setting, natural symbolic monotonicity properties arise: for example, the ECN marking probability should increase as queue length, delay, or drop rate increase. Similarly, robustness properties are desirable to prevent small fluctuations in traffic measurements from causing large oscillations in marking behavior. These properties can be directly encoded using \sys and analyzed with existing verification engines.

A second example is FIRM~\cite{qiu2020firm}, a DRL agent to dynamically adjust resource allocations across CPU, memory, cache, disk, and network bandwidth to prevent SLA violations in microservices. The agent’s input features include workload characteristics, resource utilization, and SLA satisfaction metrics.
Here, symbolic monotonicity properties naturally express expectations such as allocating fewer resources as SLA satisfaction improves or resource utilization decreases, while robustness properties capture stability under small measurement noise. These properties can be encoded using \sys and, given FIRM’s relatively small network size, we expect them to be tractable to analyze using existing verification engines.

\isignpost{Other DNN architectures} Consistent with prior verification-based studies in this domain~\cite{eliyahu2021verifying, dethise2021analyzing}, our evaluation focuses on policy networks with fully connected layers and ReLU activations.
Nevertheless, our symbolic property formulation and comparative encoding are architecture-agnostic: they rely only on comparing the outputs of two related executions of the same policy.
As such, the same properties apply to agents using other architectures or activations (e.g., RNNs, tanh, LeakyReLU), which appear in some systems and networking DRL agents~\cite{dery2023queuepilot}.
Extending analysis to these models primarily depends on backend solver support, which continues to improve (e.g., see recent developments on Alpha-Beta-CROWN~\cite{shi2025neural}).

\isignpost{Verification-aware DRL design and property enforcement}
As demonstrated by our case studies, the architecture of a DRL agent's DNN impacts its verifiability, particularly choices such as activation functions and network size. When these choices do not compromise performance, using piecewise-linear activations and more compact networks can substantially simplify symbolic analysis.
Moreover, analysis of symbolic properties can help inform the enforcement of desired behaviors.
When violations are discovered, the resulting counterexamples can be incorporated into training to reinforce expected behavior~\cite{boetius2024counterexample, gangopadhyay2023counterexample}. 
Alternatively, symbolic property compliance can potentially be used as an auxiliary cost metric in a constrained DRL formulation. Finally, monotonicity can be enforced directly through architectural constraints on the policy network, leveraging recent advances in monotonic neural networks from the supervised learning literature~\cite{runje2023constrained, wehenkel2019unconstrained}.

\isignpost{Probabilistic action selection}
DRL agents can use a $\textit{softmax}$ layer to convert output logits into a probability distribution over discrete actions, enabling stochastic exploration and uncertainty-aware decision making. However, the $\textit{softmax}$ function is neither linear nor piecewise linear, making it difficult to encode directly using existing DNN verification techniques.
In settings where the final action is selected deterministically via an $argmax$, this challenge can be avoided by reasoning directly about the ordering of logits, as we do in this work. However, omitting $\textit{softmax}$ precludes reasoning about action probabilities themselves, which is useful for agents that \emph{select one of the discrete actions probabilistically}.
Extending our approach to such cases is an important open direction, and recent work on probabilistic and convex relaxations of $\textit{softmax}$-based policies offers promising building blocks toward this goal~\cite{wei2023convex}.

\isignpost{Non-numerical action spaces}
Most DRL agents used in systems and networking operate over \emph{numerical action spaces}, where actions correspond to ordered control values such as rates, resource allocations, or thresholds, and naturally support comparisons and trends such as monotonicity and bounded change~\cite{mao2017neural, jia2024dancing, yan2021acc, wang2024autothrottle, zangooei2023flexible, qiu2020firm, tang2022abs, dery2023queuepilot}.
Extending our approach to DRL agents with non-numerical or unordered action spaces -- such as categorical decisions without an inherent ordering~\cite{bojja2019learning, hu2023giph, mao2019learning} -- is an interesting direction for future work.

\section{Related Work}

\isignpost{Verifying DRL agents in systems and networking}
Several recent works have explored using general-purpose MIP solvers and DNN verification engines to analyze DRL-based control policies in systems and networking. WhiRL~\cite{eliyahu2021verifying} uses the Marabou SMT-based verifier to check safety and liveness properties of Pensieve under \emph{fixed, extreme input conditions}, such as excellent or worst-case network states. Similarly, Dethise et al.~\cite{dethise2021analyzing} and UINT~\cite{huang2023toward} encode DRL policies as MIP or SMT problems to verify \emph{local robustness properties} around concrete input-output pairs.
These approaches demonstrate the feasibility of applying formal verification tools to DRL agents. However, they are inherently limited to \emph{point-based} or local properties defined around specific inputs, which provide limited coverage of the agent’s operational input space (see Fig.~\ref{fig:symbolic}). Our work enables analyzing symbolic properties defined over ranges of inputs rather than individual points.

\isignpost{Empirical analysis and interpretability without formal guarantees}
A complementary line of work focuses on understanding or stress-testing DRL agents through empirical or interpretability-based techniques.
Metis~\cite{meng2020interpreting} approximates DRL policies with decision trees to derive human-interpretable rules, but at the cost of reduced faithfulness to the original model; for instance, the authors report a faithfulness of only $84\%$ with respect to the original DNN.
Other approaches use DRL to synthesize network conditions under which a given algorithm underperforms~\cite{gilad2019robustifying}, or use active learning to automate the performance evaluation of congestion control schemes~\cite{pazhooheshy2023harnessing}. 
In~\cite{chakravarthy2022property}, the authors propose a robustness metric for a DRL-based controller, defined as the ratio of states where the DNN is locally robust to the total number of sampled states. Finally, Dethise et al.~\cite{dethise2019cracking} apply interpretability tools to analyze model inputs and identify anomalous or undesirable behaviors.
While effective for uncovering performance or effectiveness issues, they do not provide formal guarantees of safety or correctness of input regions.

\isignpost{Robustness via training-time regularization}
Shen et al.~\cite{shen2020deep} propose a training-time regularization technique that encourages smoother DRL policies by penalizing differences between actions taken under nominal and perturbed state inputs. Such robustness-oriented regularization can empirically reduce sensitivity to input noise, but it does not provide formal guarantees about the resulting policy's behavior. Our verification-based approach is complementary: rather than modifying the training objective, it enables post-training, formal assessment of whether a learned policy satisfies robustness and other safety properties over specified input regions.

\isignpost{Sensitivity analysis using Lipschitz-based methods}
The Lipschitz constant has been proposed as a measure of neural network sensitivity~\cite{virmaux2018lipschitz}. Given a function $f$, it bounds the maximum change in the norm of the output vector relative to changes in the norm of the input. When $f$ has a multi-dimensional output, this bound is defined over norms of the full output vector and does not track changes in the induced decision rule, e.g., the $argmax$ over action logits, which determines the DRL agent's action.
Methods such as AutoLip~\cite{virmaux2018lipschitz} compute upper bounds on the Lipschitz constant of a DNN using Jacobian-based analysis, and are therefore well-suited for reasoning about numerical output sensitivity. However, for DRL agents with discrete action spaces, where actions are selected via an $argmax$ over output neurons, small changes in the logits -- well within a Lipschitz bound -- can still lead to different action selections. As a result, global Lipschitz bounds on network outputs do not directly characterize the stability of the resulting control decisions in DRL agents considered by this work.

\isignpost{Formal methods in systems and networking}
Formal methods have been extensively applied to non-ML-based systems and networking algorithms~\cite{arun2021toward, agarwal2024towards, arashloo2023formal, liu24kivi, sun24anvil}. 
Our work contributes to the emerging effort~\cite{eliyahu2021verifying, dethise2021analyzing, wu2022scalable} to bring similar rigor to DRL-based systems, complementing prior verification efforts while expanding their applicability to symbolic, range-based properties.

\section{Conclusion}

Deep reinforcement learning is increasingly used in control loops in systems and networking, yet reasoning about agent behavior beyond individual input points remains difficult. In this work, we studied symbolic properties that capture expected behaviors over ranges of system states and showed how they can be analyzed using existing DNN verification engines via comparative encoding and decomposition.
Through an extensive empirical study using our framework, \sys, across adaptive video streaming, wireless resource allocation, and congestion control, we demonstrated that symbolic analysis substantially broadens coverage beyond point-based checks, uncovers non-obvious counterexamples, and exposes practical trade-offs related to training, model size, and solver choice. These results clarify both the promise and practical scope of symbolic property analysis for DRL agents in systems and networking.

\section*{Acknowledgments}
We would like to thank the anonymous reviewers for their valuable feedback. This work was supported in part by a Canada Research Chair grant CRC-2023-00035, an NSERC Discovery grant RGPIN-2023-03775, the Rogers Communications Chair in Network Automation, the Canada Research Chair in Network Intelligence, NSERC Alliance, Mitacs, and the Ontario Research Fund – Research Excellence program (Project \#ORF-RE012-051) from the Province of Ontario. The work has also received funding from the European Union’s Horizon Europe research and innovation programme under the ENVELOPE project (Grant Agreement No.~101139048). This work was supported in part by the DFG grant CLAYRE (565652476). The views expressed herein are those of the authors and do not necessarily reflect those of the Province.

\bibliographystyle{ACM-Reference-Format}
\bibliography{acmart}

\received{January 2026}
\received[revised]{March 2026}
\received[accepted]{April 2026}

\appendix
\section{Probability Calculation Details for the Aurora Case Study} \label{app:probab}

Let $Y_1 \sim \mathcal N(L^1_0, \sigma^2)$ and $Y_2 \sim \mathcal N(L^2_0,\sigma^2)$ represent the selected action of each policy. Consider the constraints $(L^1_0\ge \mu)$ and $(L^2_0\le -\mu)$ derived from the first property for Aurora in \S\ref{sec:aurora}.
Under these constraints, we can calculate the probability of getting a positive action $y_1$ from the first policy copy and a negative action $y_2$ from the second policy copy in the following way.

The ``hardest'' case under those constraints is at the boundary $(L^1_0=\mu,\ L^2_0=-\mu)$. Then, $\Pr(Y_1>0)=\Phi\!\left(\frac{\mu}{\sigma}\right), \text{ and } \Pr(Y_2<0)=\Phi\!\left(\frac{\mu}{\sigma}\right)$, where $\Phi(\cdot)$ is the standard normal Cumulative Distribution Function (CDF). For both events to hold jointly, assuming independence, $\Pr(Y_1>0,\ Y_2<0) = \Pr(Y_1>0)\Pr(Y_2<0) = \Phi\!\left(\frac{\mu}{\sigma}\right)^2$.

On the other hand, to ensure with probability $Q$ that we get a positive action $y_1$ from the first policy copy and a negative action $y_2$ from the second policy copy by putting the constraints $(L^1_0\ge \mu)$ and $(L^2_0\le -\mu)$, we should select $\mu = \sigma\,\Phi^{-1}\!\big(\sqrt{Q}\big)$.

\end{document}